\documentclass[fleqn,usenatbib]{mnras}

\usepackage{newtxtext,newtxmath}
\usepackage[T1]{fontenc}
\usepackage{ae,aecompl}

\usepackage{multirow}
\usepackage{graphicx}   
\usepackage{amsmath}    
\usepackage{amssymb}    
\usepackage{color}

\newcommand\code[1]{\textsc{\MakeLowercase{#1}}}
\def\zsun{{Z\rm}_{\odot}}
\def\msun{{M\rm}_{\odot}}
\def\lsun{{L\rm}_{\odot}}
\def\msunyr{{M\rm}_{\odot}/{\rm yr}}
\def\fd{f_{\rm d}}

\def\myr{{\rm Myr}}
\def\mum{\mu{\rm m}}

\def\simgt{\lower.5ex\hbox{\gtsima}} 
\def\simlt{\lower.5ex\hbox{\ltsima}}

\definecolor{bittersweet}{rgb}{1.0, 0.44, 0.37}

\newcommand{\quotes}[1]{``#1''}

\defcitealias{Laporte2017}{L17}
\defcitealias{pallottini_althaea}{P17}
\defcitealias{Ferrara2016}{F16}
\def\P17cit{\citetalias{pallottini_althaea}}
\def\L17cit{\citetalias{Laporte2017}}
\def\F16cit{\citetalias{Ferrara2016}}

\title[Dusty galaxies in the EoR]{Dusty galaxies in the Epoch of Reionization: simulations}

\author[C. Behrens et al.]{
C. Behrens,$^{1}$\thanks{\href{mailto:christoph.behrens@sns.it}{christoph.behrens@sns.it}},
A. Pallottini,$^{1,2,3,4}$,
A. Ferrara$^{1,5}$,
S. Gallerani$^{1}$, 
L. Vallini$^{6}$\\
$^{1}$Scuola Normale Superiore, Piazza dei Cavalieri 7, I-56126 Pisa, Italy\\
$^{2}$Centro Fermi, Museo Storico della Fisica e Centro Studi e Ricerche ``Enrico Fermi'', Piazza del Viminale 1, Roma, 00184, Italy\\
$^{3}$Cavendish Laboratory, University of Cambridge, 19 J. J. Thomson Ave., Cambridge CB3 0HE, UK\\
$^{4}$Kavli Institute for Cosmology, University of Cambridge, Madingley Road, Cambridge CB3 0HA, UK\\
$^{5}$Kavli IPMU, The University of Tokyo, 5-1-5 Kashiwanoha, Kashiwa 277-8583, Japan\\
$^{6}$Nordita, KTH Royal Institute of Technology and Stockholm University, Roslagstullsbacken 23, SE-10691 Stockholm, Sweden
}

\date{Accepted XXX. Received YYY; in original form ZZZ}
\pubyear{2017}

\begin{document}
\label{firstpage}
\pagerange{\pageref{firstpage}--\pageref{lastpage}}
\maketitle

\begin{abstract}
The recent discovery of dusty galaxies well into the Epoch of Reionization (redshift $z>6$) poses challenging questions about the properties of the interstellar medium in these pristine systems. By combining state-of-the-art hydrodynamic and dust radiative transfer simulations, we address these questions focusing on the recently discovered dusty galaxy A2744\_YD4 ($z=8.38$, \citealt{Laporte2017}).
We show that we can reproduce the observed spectral energy distribution (SED) only using different physical values with respect to the inferred ones by \cite{Laporte2017}, i.e. a star formation rate of SFR = 78 $M_\odot \rm yr^{-1}$, a factor $\approx 4$ higher than deduced from simple Spectral Energy Distribution fitting. In this case we find: (a) dust attenuation (corresponding to $\tau_V=1.4$) is consistent with a Milky Way extinction curve; (b) the dust-to-metal ratio is low, $f_\mathrm{d} \sim 0.08$, implying that early dust formation is rather inefficient; (c) the luminosity-weighted dust temperature is high, $T_d=91\pm 23\, \rm K$, as a result of the intense ($\approx 100\times$ MW) interstellar radiation field; (d) due to the high $T_d$, the ALMA Band 7 detection can be explained by a limited dust mass, $M_d=1.6\times 10^6 \msun$.  Finally, the high dust temperatures might solve the puzzling low infrared excess recently deduced for high-$z$ galaxies from the IRX-$\beta$ relation. 
\end{abstract}


\section{Introduction}
Understanding the formation and evolution of the first galaxies lighting up at the very end of the cosmic dark ages is among the key challenges in modern astrophysics.
Studying the properties of such primeval objects and their interaction with the cosmic environment is crucial not only for the field of galaxy formation, but it also connects the small-scale physics of the interstellar medium, stellar evolution, and feedback processes to the cosmological scales.
The appearance of these galaxies provided a significant fraction to the budget of ionizing photons that drove the Epoch of Reionization \citep{Madau1999,Gnedin2000,Barkana2001,Sokasian2003,Mesinger2007}, the last phase transition of the Universe, and the metals and dust produced in these objects were the first pollutants of the pristine circumgalactic/intergalactic medium, laying out the substrate that fed later stages of structure formation \citep{Madau2001,Ferrara2008,Meiksin2009,Pallottini2014,Ferrara2016a,Jaacks2017}.

Many open questions about the properties of the first galaxies remain to be solved: Which stellar populations do they host? What processes control their star formation activity? How do they affect their intergalactic environment?
A specific and interesting set of questions concerns the nature of their metal and dust content, in particular how efficient the production of dust is in these extreme objects, given their age, high specific star formation rates, and turbulent environments.

Only the latest enhancements in our observational capabilities allow to detect and characterize objects at these high redshifts \citep{Maiolino2015,Willott2015,Capak2015,Jiang2016,Knudsen2016,Bouwens2016,Pentericci2016,Carniani2017}, with near-future instruments like JWST, SKA, SPICA \citep{Gardner2006,Taylor2013,spinoglio:2017,egami:2017} providing great prospects of improving our view on this epoch.

The question for the dust content of the first galaxies is not answered, yet.
For example, the sample of \cite{Capak2015} is dominated by low-dust, UV-bright galaxies.  
On the other hand, \cite{Watson2015} report on a detection at $z=7.5$ with a dust mass of several times 10$^7$ M$_\odot$, at a relatively low stellar mass of 10$^9$ M$_\odot$.

In particular, recently, \citet[][\L17cit hereafter]{Laporte2017}  reported the detection of a very young galaxy (A2744\_YD4), lensed by Abell 2744, that is located at the fringe of Cosmic Dawn, with a redshift of $z=8.38$. Using multi-wavelength data from HST, XSHOOTER, and ALMA, they measure its SED from the rest frame UV to the far Infrared (FIR). They estimate the stellar/dust mass, star formation rate (SFR), and other properties using both simple analytic estimates and SED fitting, and find A2744\_YD4 to have a stellar mass of around 10$^9$ M$_\odot$, a SFR of about 20 M$_\odot$/yr, and a dust mass of $\sim 6 \times 10^6$ M$_\odot$.

To improve our understanding, we can model this problem with numerical simulations. In particular, zoom-in simulations of high-redshift galaxies can be used to capture the formation and evolution of the structure of galaxies in their cosmological environment by consistently following the star formation history and feedback processes that drive the chemical evolution of galaxies at high-z \citep[e.g.][]{hopkins:2017,Pallottini2016,fiacconi:2017mnras,ceverino:2017}. Then, such hydrodynamical simulation can be post-processed with in-depth radiative transfer calculations to understand the influence of geometry, composition, dust mass and distribution, and to compare with what is observed. For low redshift, e.g. \cite{Camps2016} have done such radiative transfer simulations on top of the EAGLE simulation suite \citep{McAlpine2016}. Additionally, \citet{Safarzadeh2017} have performed radiative transfer at intermediate redshifts, and recently, \cite{Narayanan2017} have investigated the IRX-$\beta$ relation for high-redshift galaxies by running radiative transfer for a series of high-resolution galaxy simulations; complementary, \cite{Popping2017} present an analytic model to understand how stellar age, turbulence, and geometry influence the IRX-$\beta$ relation.

The simulations of \citet[][see Sec. \ref{sec_hydro_sim}]{pallottini_althaea} follow the evolution of a high-redshift galaxy named \quotes{Alth{\ae}a}, with a mass and SFR similar to  A2744\_YD4 at the relevant redshift. It is therefore natural to ask whether the observational signatures to be expected from Alth{\ae}a are similar to what is observed by \L17cit.

The absorption of star light by dust and the subsequent reemission cause the transfer of energy from the rest frame optical/UV to the IR. Thus, such observations encode not only information about the stellar emission, but also about the dust mass and distribution: this also allows us to put constraints on the metal and dust properties of high-redshift galaxies, in particular about the fraction of metals locked up in dust and its composition.

In this paper, we therefore focus on radiative transfer calculations as post-processing step on Alth{\ae}a. We employ a publicly available code (\code{SKIRT}) and perform simulations of attenuated stellar emission over a broad range of wavelengths and the subsequent reemission of energy by the heated dust grains. Our goal is to analyse whether or not synthetic fluxes in Alth{\ae}a are compatible with the data from \L17cit, to understand what the important physical parameters are, and to which extent we can constrain them from comparing the observation to the theoretical model.

Since sophisticated simulations like Alth{\ae}a are still computationally very demanding given the complexity of the physics involved, it is currently not feasible to produce a statistically significant sample of such objects \citep[see][]{ceverino:2017}. Owing to these limitations, we present our work here using a number of snapshot of one specific simulation as a proof-of-concept to show that this type of simulations yield useful insights for the understanding of high-redshift objects. We cannot exclude that A2744\_YD4 differs significantly from our simulated galaxy in its evolution, morphology, and kinematics. However, \cite{pallottini_althaea} have shown that Alth{\ae}a can reproduce many observational properties of high-z star forming galaxies, e.g. the SFR-$M_{\star}$, the Schmidt-Kennicutt, and the CII-SFR relations (see Fig. 3, 9, and 12, respectively therein). If A2744\_YD4 can be viewed as a typical star-forming galaxy as well, a comparison between the two seems more than fair.

The paper is organized as follows: In Sec. \ref{sec_model}, we provide a description of both the hydrodynamical simulation that is the basis for our calculation (Sec. \ref{sec_hydro_sim}), and the model choices we made for the radiative transport simulations (Sec. \ref{sec_rt_dust}). In Sec. \ref{sec_results}, we present our results and compare to the \L17cit data; in particular we discuss how the gas morphology affects the UV and IR emission, and the implication for the IRX-$\beta$ correlation. In Sec. \ref{sec_sed_fitting} we address to which extent SED fitting can recover our results. Finally discussion and conclusions are summarized in Sec. \ref{sec_conclusions}.


\section{Model}\label{sec_model}
In the following, we briefly describe the hydro simulation in Sec. \ref{sec_hydro_sim}, and the different parts of the radiation transfer model in Sec. \ref{sec_rt_dust}.
We stress that we do the RT as a post-processing step on snapshots of the hydro simulation, since with the current state of the art it is unfeasible for a fully coupled simulation to follow the evolution of the system with a precision high enough to have a fair comparison with observations.

\subsection{High-$z$ galaxy simulation}\label{sec_hydro_sim}

The hydrodynamical simulation used in this work is fully described in \citet[][hereafter \P17cit]{pallottini_althaea} and is summarized as follows.

\P17cit use a modified version of the adaptive mesh refinement code \code{RAMSES} \citep{teyssier:2002} to zoom-in on the evolution of a $z\sim 6$ dark matter (DM) halo of mass $\sim 10^{11} \msun$ from cosmological initial condition. The gas has mass resolution of $10^4 \msun$, and is resolved to spatial scales of $\simeq 30\,{\rm pc}$.

Stars are formed from molecular hydrogen, the abundance of which is calculated by a chemical network implemented via the \code{KROME} code \citep{grassi:2014mnras,bovino:2016aa}.
The thermal and turbulent energy content of the gas is modelled according to \citet[][]{agertz:2015apj}. As detailed in \citet{Pallottini2016}, stellar feedback includes supernovae, winds from massive stars and radiation pressure; the stellar energy inputs and chemical yields depend stellar populations and age.
In particular, age-dependent chemical yields are calculated from stellar evolutionary models, by interpolating {\tt padova} stellar tracks for $Z_{\star}/\zsun = 0.02,\, 0.2,\, 0.4,{\rm and}\, 1$ \citep{padova:1994} and by adopting a \citet{kroupa:2001} initial mass function \citep[see][in particular Fig. 2]{Pallottini2016}.
Additionally, the model accounts for the unresolved physics inside the molecular clouds for the blastwave propagation. Currently, it does not include full radiative feedback, in particular photoevaporation of molecular clouds. Such processes might change the structure of the interstellar medium (ISM) by shortening the lifetime of molecular clouds \citep{Decataldo2017}.

The selected DM halo hosts Alth{\ae}a, a prototypical Lyman Break Galaxy (LBG) that has a stellar mass $M_\star\sim 10^{10} \msun$ and a $\rm SFR\sim 100\, \msun {\rm yr}^{-1}$ at $z\sim6$.
Along its evolution, Alth{\ae}a SFR-stellar mass relation is compatible with high-$z$ observations \citep[e.g.][]{Jiang2016}, and agrees with the Schmidt-Kennicutt relation \citep{Schmidt1959,Kennicutt1998,kennicutt:2012,krumholz:2012apj}.

\subsection{Radiation transfer in a dusty medium}\label{sec_rt_dust}

Continuum radiative transfer in a dusty medium is a well-studied problem \citep[e.g.][and references therein]{Ferrara1999}, and a number of codes have been developed to tackle this problem \citep[e.g.][]{Jonsson2006,Robitaille2011} mostly relying on sophisticated Monte-Carlo methods. We use the publicly available code \code{SKIRT}\footnote{version 7.4, \url{http://www.skirt.ugent.be}} \citep{Baes2003,Baes2015,Camps2015,Camps2017}, to post-process the output of our hydrodynamical simulations at different ages of the simulated galaxy, that is, for different snapshots.

The general code pipeline can be summarized as follows. Given a simulated galaxy, \code{SKIRT} solves the radiative transfer problem in two steps.

Firstly, given the location of stellar emission \code{SKIRT} launches tracer photons, sampling a given intrinsic SED for the stellar continuum emission (Sec. \ref{sec_stellar_emission}).
Photon packages are absorbed and/or scattered as they probe the dust distribution in the simulation volume (Sec. \ref{sec_model_dust}), according to the dust properties and composition (Sec. \ref{sec_model_dust_2}). The amount of energy absorbed by each of the dust cells in the simulations is kept track of. This first step yields the attenuated SED of the stellar emission, after being processed by the dusty ISM.
In a second step, the energy absorbed by dust is reemitted at infrared wavelengths. This requires an additional model for connecting the amount of energy absorbed in each cell to its wavelength-dependent emissivity (Sec. \ref{sec_model_dust_3}). Similarly to the first step, tracer photons are launched from dust cells, sampling the spatial/frequency distribution of the emitting dust. 

Both steps together yield a self-consistent solution to the radiative transfer problem, ranging from the far-UV to the far-IR.

\subsubsection{Stellar emission}\label{sec_stellar_emission}
Location of the stellar emission is given by positions of stellar particles in the hydrodynamical simulation. \code{SKIRT} provides the possibility to directly import a file containing the relevant information.

Using \code{SKIRT}, we build the SEDs of individual stellar particles from the \cite{Bruzual2003} family of stellar synthesis model, given the mass, age, and metallicity of each star cluster, effectively treating each as a single starburst. This data is readily available from our hydro simulations.

The SED (for both stellar and dust emission) is sampled using 150 logarithmically spaced bins in wavelength, covering 0.1 to 1000 $\mu{\rm m}$. To ensure a good convergence of the results, a total of 10$^7$ photon packages per wavelength bin is launched from the stars at each calculation. For comparison, \cite{Camps2016} used $10^6$ photons per wavelength bin and a higher number of wavelength bins (450). Note that the number of bins in wavelengths does not affect our results significantly, as is shown in App. \ref{appendix_convergence}.

\subsubsection{Dust spatial distribution}\label{sec_model_dust}
The spatial distribution of dust is imported from the hydrodynamical simulation by keeping the same spatial resolution of the AMR grid. We do not consider dust in gas patches that have $Z< 10^{-3}\zsun$, i.e. the metallicity floor used as initial conditions in \P17cit to mimic high-$z$ enrichment from external galaxies \citep[]{maio:2010,Pallottini2014}.

In the galaxy simulation \P17cit follow the enrichment of metals that are produced by the star clusters. During the evolution of Alth{\ae}a, the fraction of the metal mass locked up in dust is
\begin{equation}\label{eq_def_f_d}
f_\mathrm{d} = M_\mathrm{D}/M_\mathrm{Z}\,,
\end{equation}
and it is assumed to be fixed at $f_\mathrm{d}=0.3$, one of the values commonly used for the {Milky Way (MW)}. Practically speaking, the choice of $f_\mathrm{d}$ provides a normalization to the extinction curve. The radiative transfer model is very sensitive to the value of $f_\mathrm{d}$, that is poorly constrained at high-$z$ by observations \citep[also see][and reference therein]{Wiseman2017} and models give a wide range of values \citep{asano:2013,Aoyama2017}; thus in the following we perform different radiative transfer calculations by allowing it to vary.

\subsubsection{Dust absorption and scattering}\label{sec_model_dust_2}

The size distribution and composition of the dust determine its absorption and scattering properties \citep[e.g.][]{draine:2014apj}.
We use models from \protect\cite{Weingartner2001} that are appropriate for the extinction curve of MW and Small Magellanic Cloud (SMC)\footnote{While the MW dust model is already available in the public version of \code{SKIRT}, we implemented the SMC model from \cite{Weingartner2001}.}. 
The grain size distribution of graphite grains/silicate grains/polycyclic aromatic hydrocarbons (PAHs) is sampled using 5/5/5 bins. Note that in the SMC model we effectively set the fraction of dust in PAHs to zero.
We do not consider self-absorption of dust emission; however, to check its possible influence onto the SEDs, we reran one of our simulations with self-absorption switched on, see Sec. \ref{sec_sed_fitting} and App. \ref{appendix_convergence}.

The exact nature of the dust grains in high-redshift galaxies is a matter of ongoing debate; we chose these two models for simplicity \cite[but see][and references therein]{Gallerani2010}.
Note that due to the different total dust mass and spatial distribution, the resulting attenuation curves from our simulations are not expected to be the same as for the MW/SMC (see later in Sec. \ref{sec_finding_a_fit}).

Finally, note that recently \citet{dalcanton:2015apj,planck2016XXIX} noticed a discrepancy between the dust model from \citet{draine:2014apj} and observations, reporting that the models overestimates dust masses.
Given also the high-$z$ uncertainty on dust properties, in our model we do not account for such effect; however we note that an overestimation of dust mass can be corrected by considering a lower $f_d$ (eq. \ref{eq_def_f_d}), that is a parameter of our model.

\subsubsection{Dust emission properties}\label{sec_model_dust_3}

For the reemission by dust, we assume that grains radiate as gray bodies, with an equilibrium temperature set by the balance of the incident stellar radiation field and the reemission.
Since \code{SKIRT} allows for multiple grains population, the emission from every single dust patch is a linear combination of gray bodies with different temperatures, corresponding to the different dust size distribution of the different components.

Additionally, we take into account the effect of the cosmic microwave background (CMB) on the dust temperature. As \citet{daCunha2013} point out, the CMB at these high redshifts has a temperature that can lead to an effective heating of dust, i.e. we expect a corrected temperature
\begin{equation}
T^{\rm corr}_{D}(z) = \left[T_{D}^{4+\beta_d} +T_{\rm CMB}^{4+\beta_d}(z) -T_{\rm CMB}^{4+\beta_d}(z=0)\right]^{1/(4+\beta_d)}\,,
\end{equation}
where $T_{\rm CMB}(z)= 2.725\,(1+z)\,\rm K$ is the average CMB temperature at $z$. $\beta_d$ is the spectral index of the dust emitting as a gray body.
As dust temperature approaches the CMB one, the dust emission is effectively reduced, since observations are performed against the CMB \protect\citep[see also][for the effect on FIR line emission]{Basu2007,Schleicher2008,Pallottini2015}.
Since the correction is done at runtime, we needed to choose a value for $\beta_d$ before running the simulation. We set it to $\beta_d = 2$. However, in our model $\beta_d$ is not a free parameter, as it is determined by the radiative transfer. 
A posteriori for the fiducial model (Sec. \ref{sec_finding_a_fit}) we find $\beta_d \simeq 1.7$; this discrepancy has no effect onto our results, in particular since the CMB correction has only marginal effect as we find very high dust temperatures (see Sec. \ref{sec_temperature}).

Finally, we note that \code{skirt} is capable of handling stochastic heating \citep[][]{Camps2015}, however in our model we assume the grains to be in thermal equilibrium with the incident radiation field; the reliability of our assumption is benchmarked in App. \ref{appendix_convergence}.

\section{Results}\label{sec_results}

\subsection{Spectral Energy Distribution}\label{sec_finding_a_fit}

\begin{figure}
\centering
\includegraphics[width=0.49\textwidth]{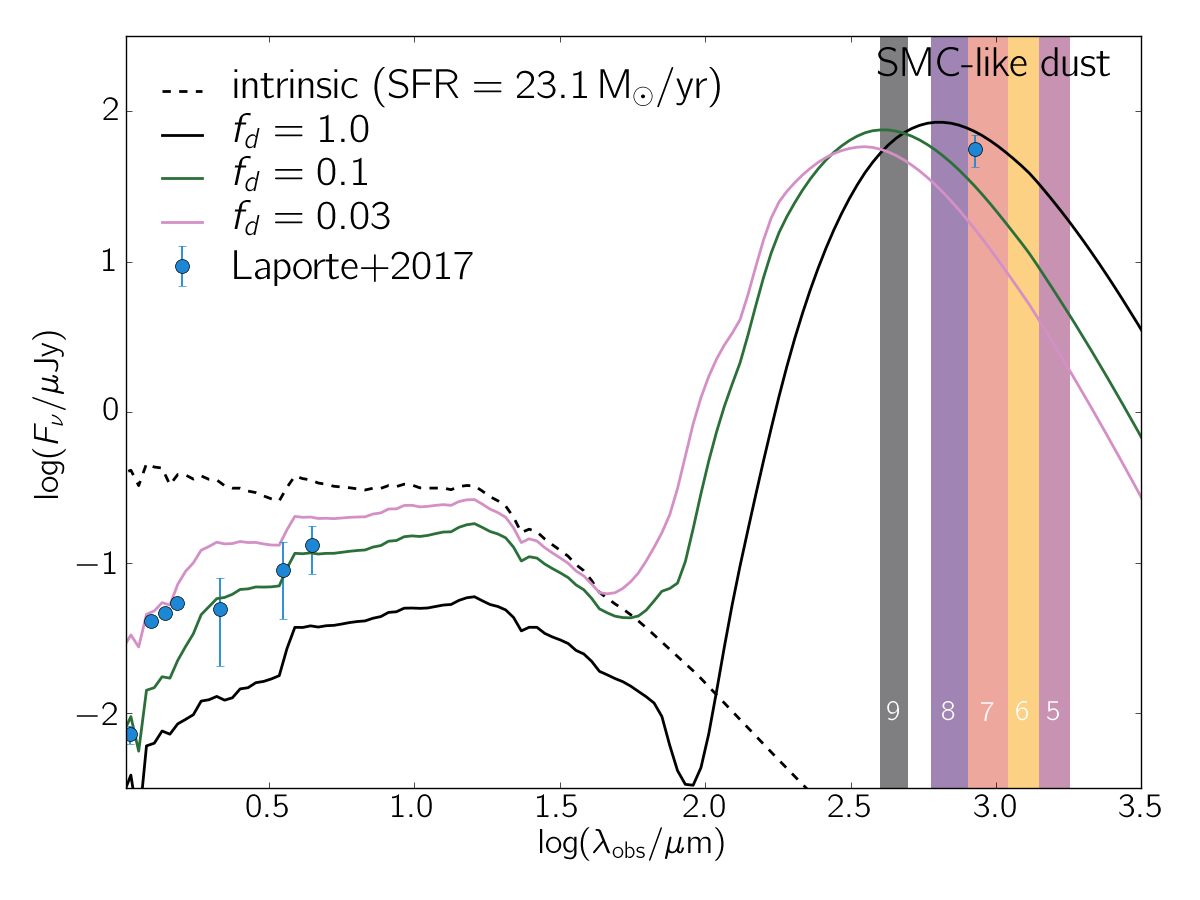}\\
\includegraphics[width=0.49\textwidth]{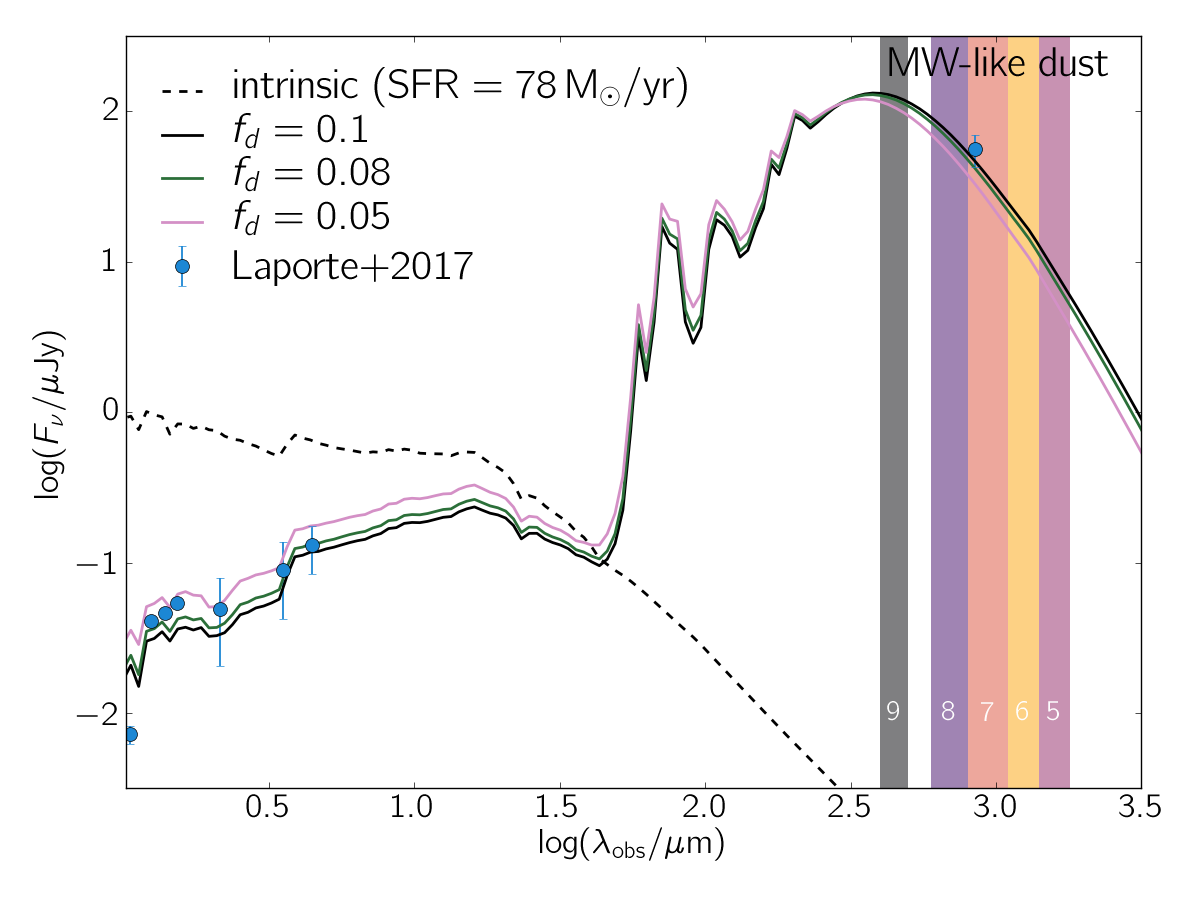}\\
\includegraphics[width=0.49\textwidth]{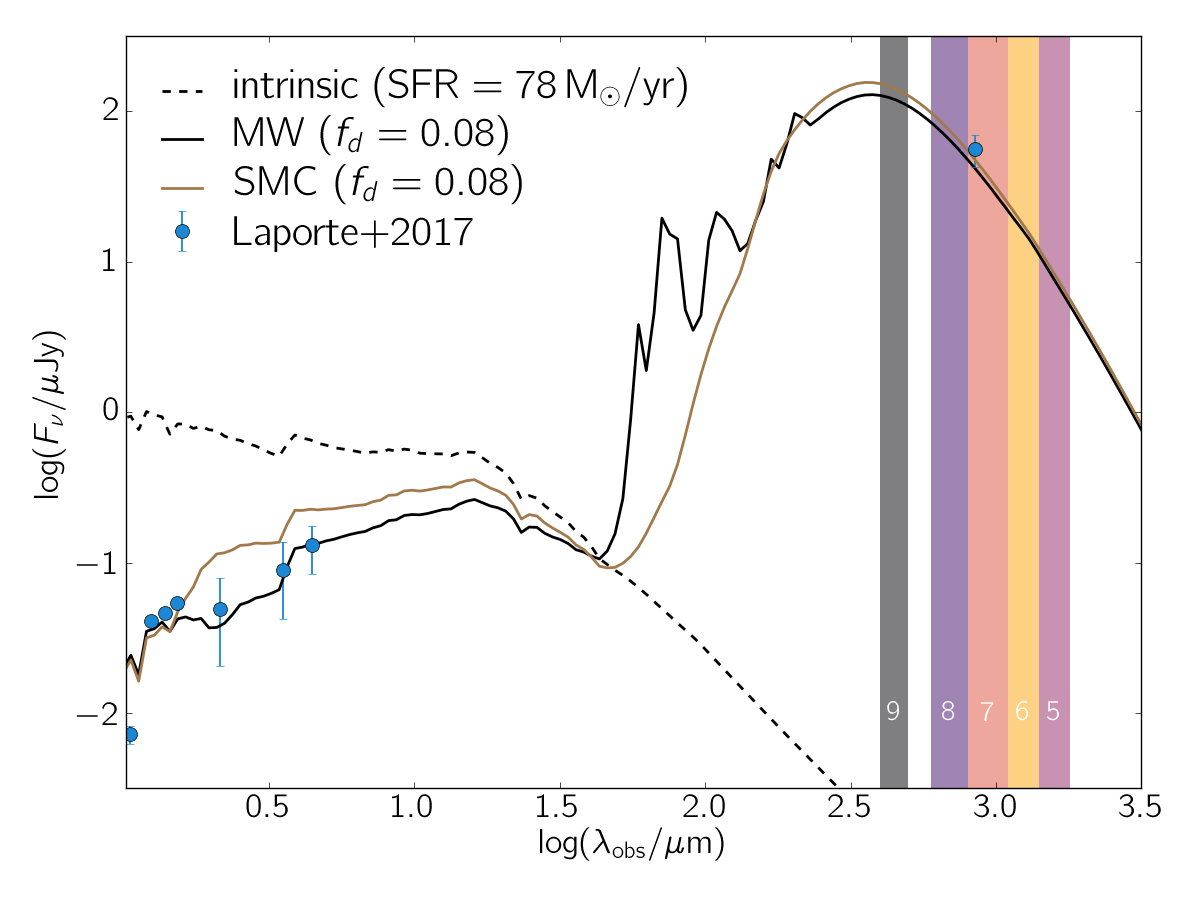}
\caption{
SED for our simulated galaxy Alth{\ae}a for different dust to mass ratio ($f_d$), dust type (SMC, MW) and galaxy star formation rate (SFR, see Tab. \ref{table_snaps}). Note the silicate absorption features in the mid-IR portion of the spectrum for the MW dust curve.
The panels highlight different variations in the parameters, as detailed in the text.
The dashed line shows the intrinsic stellar continuum from Alth{\ae}a.
The blue symbols are the observed data points from \protect\L17cit.
Shaded regions mark the ALMA bands 9 to 5.
\label{fig_sed_panels}
}
\end{figure}

\begin{table}
\centering
\begin{tabular}{cclll}
\hline
$t_{\star}$  & $M_{\star}$   & ${\rm SFR}$ & $M_{Z}$        & $z$\\
$[\myr]$     & $[10^9\msun$] & $[\msunyr]$ & $[10^6 \msun]$ & ~\\
\hline
413.3            & 5.7                     & 23.1                & 8.8  &  8.01\\
513.4            & 10.3                    & 78.0                & 12.6 &  7.2\\
\hline
\end{tabular}
\caption{Overview of the properties of the employed simulated galaxy at different ages ($t_\star$): stellar mass ($M_{\star}$), star formation rate (${\rm SFR}$), metal mass ($M_{Z}$), and redshift ($z$) during the simulation. 
\label{table_snaps}
}
\end{table}

\begin{figure}
\centering
\includegraphics[width=0.49\textwidth]{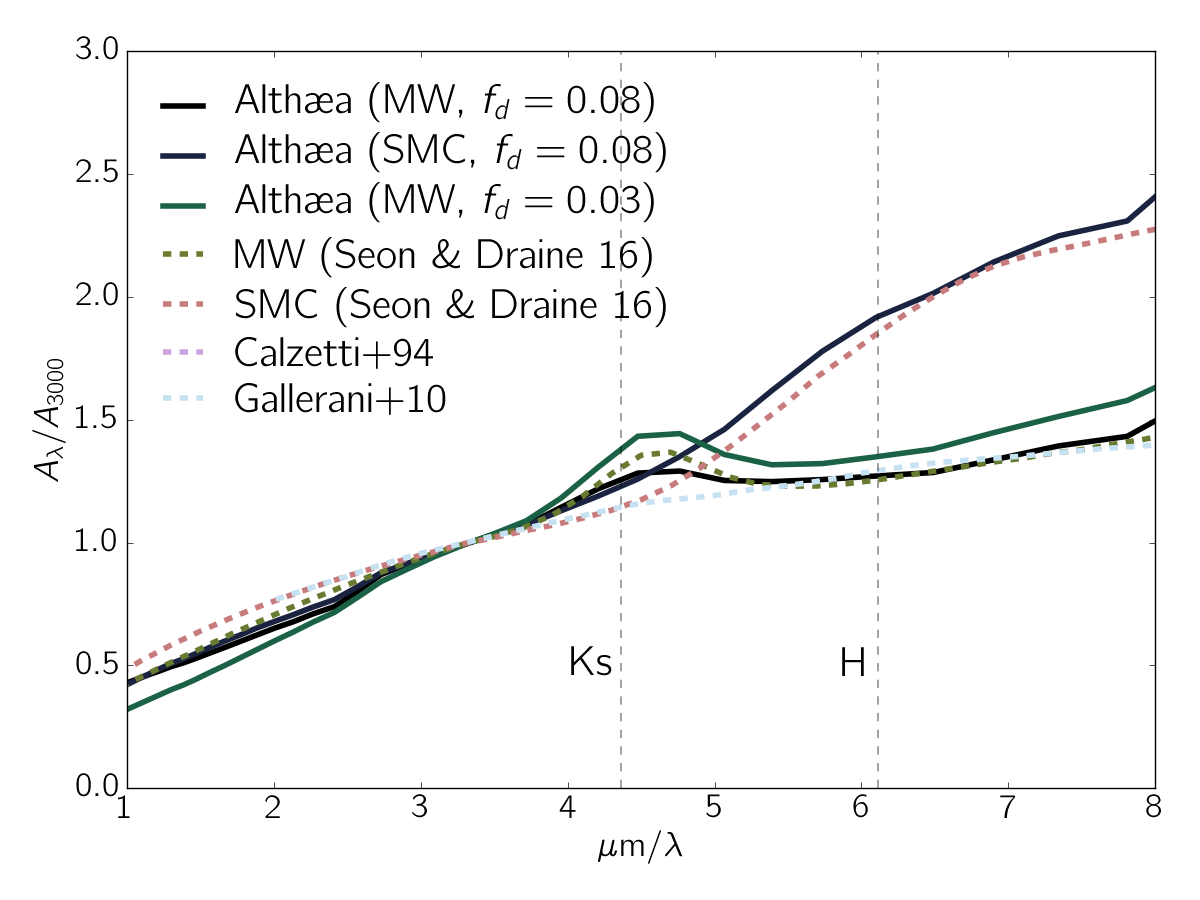}
\caption{
Attenuation curves obtained from the radiative transfer simulation (solid lines). For comparison, we show attenuation curves from Seon \& Draine (2016) for MW/SMC; we use the results for ($\tau_V$,R$_S$,M$_S$) = (8,1,20), see the paper for details. 
We also show the attenuation curve proposed by Gallerani et al. (2010) for high-redshift galaxies, and the classical Calzetti law. See the text for details.
The dotted vertical line indicate the restframe center of the Ks and H bands.
}
\label{fig_attenuation}

\end{figure}

We start by selecting a simulation snapshot close to the redshift of the \citet[][\L17cit hereafter]{Laporte2017} observed galaxy, i.e. at $z=8.01$. At this redshift, Alth{\ae}a has an age $t_{\star} = 413.3\,\myr$, a stellar mass of $M_{\star} = 5.7\times 10^9 \msun$, star formation rate ${\rm SFR} = 23.1\,\msunyr$, total dust mass $M_{D}=8.8\,\fd\, 10^6$ $M_\odot$ (see Tab. \ref{table_snaps}). These properties are in rough agreement with those inferred by \L17cit. The stellar mass is a factor of 2.8 higher in our simulation, but since the UV will be dominated by the youngest stars, the SFR is the important quantity to compare with.If not stated otherwise, all our simulation have a line of sight oriented face-on.

In the upper panel of Fig. \ref{fig_sed_panels}, we show the SED we obtain from \code{SKIRT} for a SMC-type dust for three different values of $f_\mathrm{d}$. The dashed line shows the intrinsic stellar emission, namely the SED obtained ignoring dust attenuation effects. We additionally show the observed fluxes from \L17cit (blue symbols).
As is evident from the plot, there is no good simultaneous fit of the UV and FIR data for any of these models.
The $f_\mathrm{d} = 0.1$ case reproduces well the Spitzer data points, but the spectrum falls too steeply towards the shorter wavelength HST points; in addition, this model falls short to account for the ALMA Band 7 FIR point.
Note that, if one decreases the UV attenuation by reducing the dust mass, this also decreases the amount of energy converted into IR and therefore the FIR flux. The situation is even worse for the other two more extreme cases, $f_\mathrm{d} = 0.01, 1$. We conclude that changing just the dust mass cannot resolve the discrepancy. 

To produce fluxes comparable to the observation, we found that two ingredients are needed: (a) a higher intrinsic UV luminosity, and (b) a flatter extinction curve more similar to that produced by MW-type dust grains.
The first requirement is a natural consequence of the energy balancing argument above: we need to convert enough energy to the IR, and, yet recover acceptable levels of flux in the UV; this entails a higher intrinsic UV luminosity.
To accomplish this, we looked for an evolutionary stage of Alth{\ae}a featuring a more intense star formation burst. We found one such state at a galaxy age $t_{\star} = 513.4\,\myr$ ($z=7.2$) in the simulation\footnote{Note that Alth{\ae}a has a SFR history that is exponentially increasing with age, and shows burst of width $\simeq 20\,\myr$ (see Fig. 2 in \P17cit).}. To perform a fair comparison with \L17cit observations, we shift Alth{\ae}a using a luminosity distance corresponding to $z=8.38$. We note that while the stellar mass of this snapshot is another factor of 2 higher compared to the younger snapshot owing to the accumulation of stellar mass over time, what actually drives the change in the intrinsic UV flux is the SFR, because it is the youngest stars that dominate the UV and not the bulk of the stellar mass.

We plot the SED resulting from the radiative transfer simulation on this setup using a MW-type dust in the middle panel of Fig. \ref{fig_sed_panels}. For $f_\mathrm{d} = 0.08$, we find the SED to match well UV and the IR data to be broadly consistent with observations. The above $f_\mathrm{d}$ value is small compared to the MW one ($f_\mathrm{d} \approx 0.3$). However, we note that \cite{Aoyama2017} find similar values by performing detailed hydrodynamical simulations including sub-grid models for the destruction/formation of dust \citep[see also][]{asano:2013,grassi:2017}. For example, at a galaxy age 300 Myr, they find  $f_\mathrm{d} \approx 0.1$. In the following, we will refer our run of the simulation with ${\rm SFR} = 78\,\msunyr$, $f_\mathrm{d} = 0.08$, and MW-type dust as the {\it fiducial} simulation.

As stated above, we use MW-type dust instead of SMC dust for this fiducial simulation. This is required to match the slope of the data around 1.6 $\mu$m. 
To clarify the need for MW dust, in the last panel of Fig. \ref{fig_sed_panels} we compare SEDs obtained for SMC vs MW attenuation curves. The difference in slope close to the three HST data points is clearly visible.
Inspecting more closely this issue, we find that the difference arises from the decrease of the MW extinction curve at wavelengths shorted than the 2200\AA~bump with respect to the SMC one (see Fig. \ref{fig_attenuation}. In fact, the SMC curve continues to rise towards the FUV. An SMC composition of dust would then render it very hard to explain the modest evolution of the extinction level between the Ks and the WFC3 filters seen in the data of \L17cit, while retaining the UV/IR levels at the same time. We note that the 2200\AA\, bump is visible in our simulations for small values of $f_\mathrm{d} < 0.05$. For higher values, the increased dust mass renders scatterings more important, washing out the feature, an effect also found by \cite{Ferrara1999}. For comparison, we also show attenuation curves from simulations by \cite{Seon2016} in Fig. \ref{fig_attenuation}. Their simulation dealt with a single turbulent cloud, which makes the results hard to compare; however, if we assume dust and stars to be well mixed ($R_S=1$), the attenuation curves are qualitatively similar for a relatively high optical depth ($\tau_V=8$, see also Fig. \ref{fig_map_tau}) and Mach number close to the mean one for Alth{\ae}a (M$_S=20$), as implied by \citealt{Vallini2017}, see in particular Fig. 7 therein). We also show the proposed attenuation curve from \cite{Gallerani2010} for high-redshift galaxies, and the Calzetti law \citep{Calzetti1994}.

For the remainder of the paper, we will focus on the fiducial simulation. We note that we could also vary the inclination systematically, since the viewing angle is expected to change the UV flux. We provide a basic assessment of the effect in App. \ref{appendix_convergence}.

\subsection{Morphological overview of the emission}\label{sec_overview_emission}
 
\begin{figure*}
\centering
\includegraphics[width=0.32\textwidth]{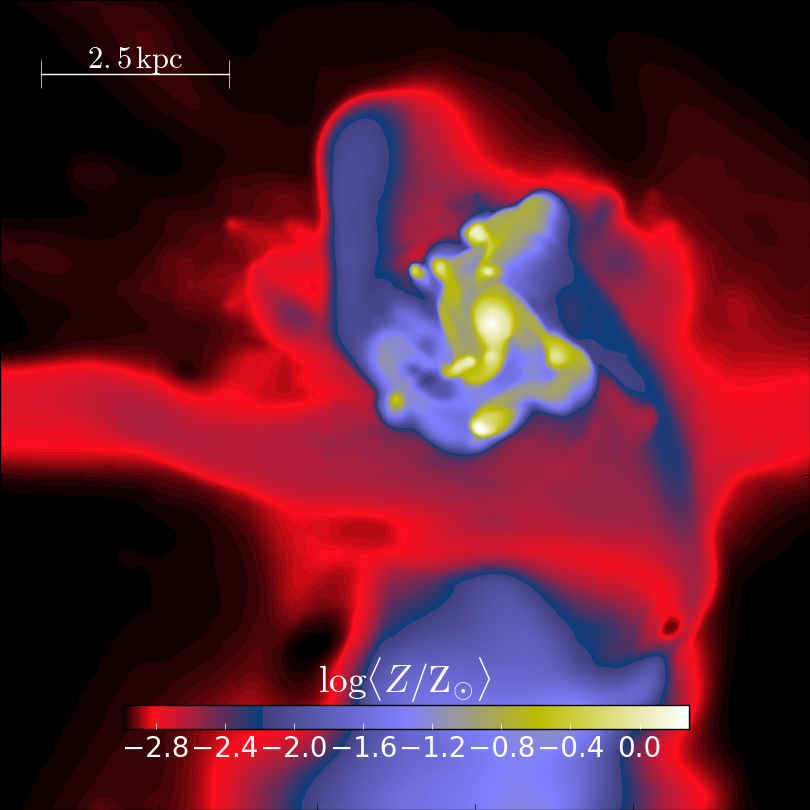}
\includegraphics[width=0.32\textwidth]{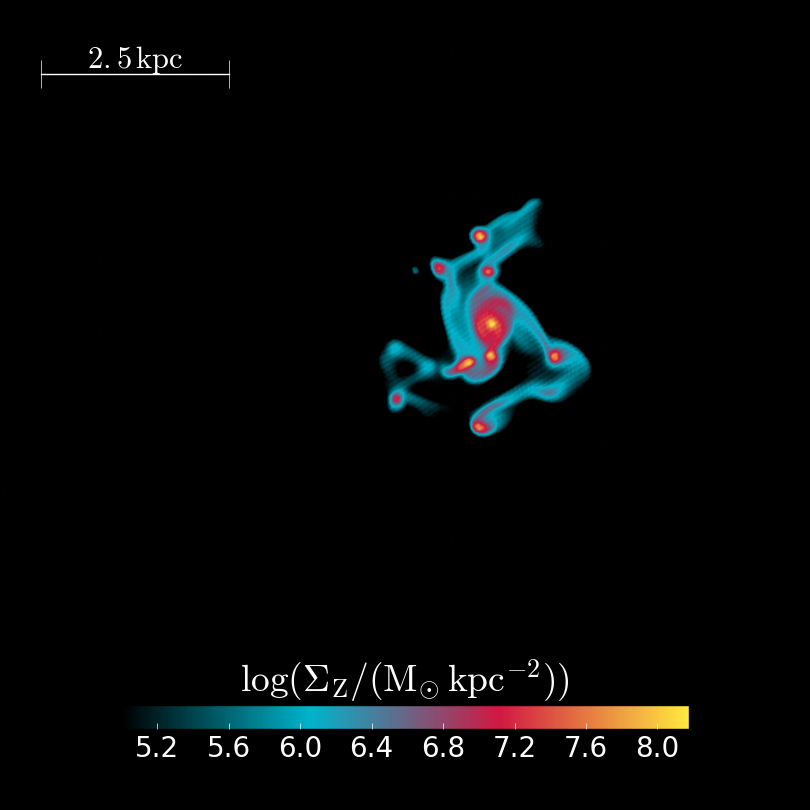}
\includegraphics[width=0.32\textwidth]{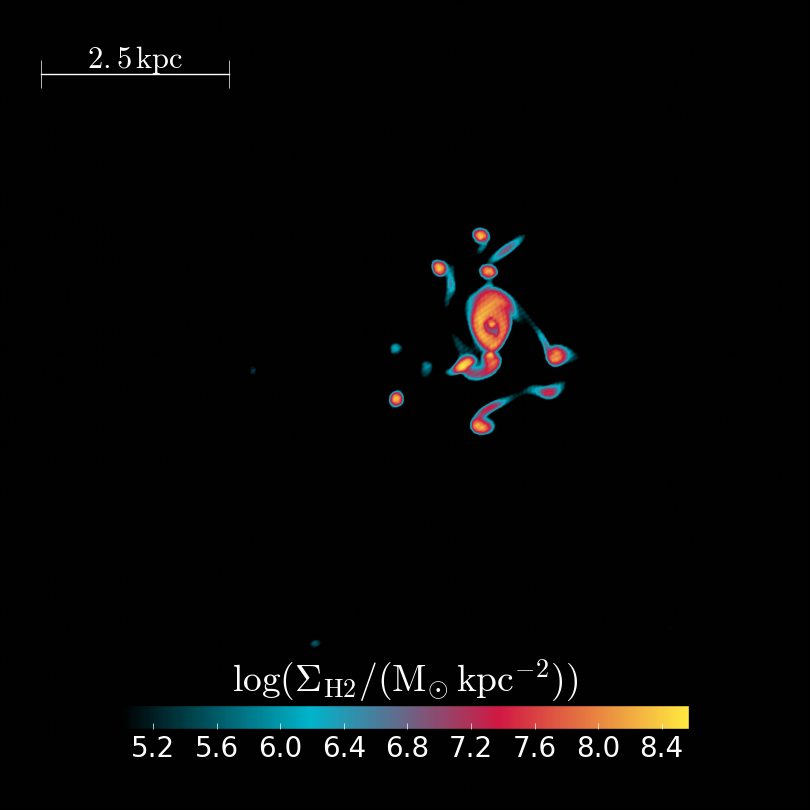}
\caption{
Face-on map of physical characteristics of Alth{\ae}a when the galaxy has ${\rm SFR} =78\, \msunyr$ (see Tab. \ref{table_snaps}).
We show the metallicity map ($Z$, left), the metal surface density ($\Sigma_Z = Z\Sigma_g$, center) and the molecular hydrogen surface density ($\Sigma_{\rm H2}$, right).
\label{fig_althaea_ph}
}
\end{figure*}

In Fig. \ref{fig_althaea_ph} we show some of the physical characteristics of Alth{\ae}a for our fiducial snapshot.
From left to right we plot the metallicity map ($Z$), the surface density of metals ($\Sigma_Z = Z \Sigma_g$) and the molecular hydrogen surface density ($\Sigma_{\rm H2}$).
We can see (left panel) that the metal pollution is extended up to about $3\,\rm kpc$ from the center of the galaxy, and there is a sharp metallicity gradient, going from $Z\simeq \zsun$ to $Z\simeq 10^{-2}\zsun$. Most of the metal (and consequently dust) mass is concentrated near the galactic disk ($r\simeq 0.5\,\,kpc$, central panel); such metals are mostly distributed in clumps, co-located with the star forming regions; such regions can be tracked by the presence of $\rm H_2$ clumps (right panel). High gas surface density are needed to self-shield the $\rm H_2$, that can efficiently form via interaction with dust grains.

\begin{figure*}
\centering
\includegraphics[width=0.49\textwidth]{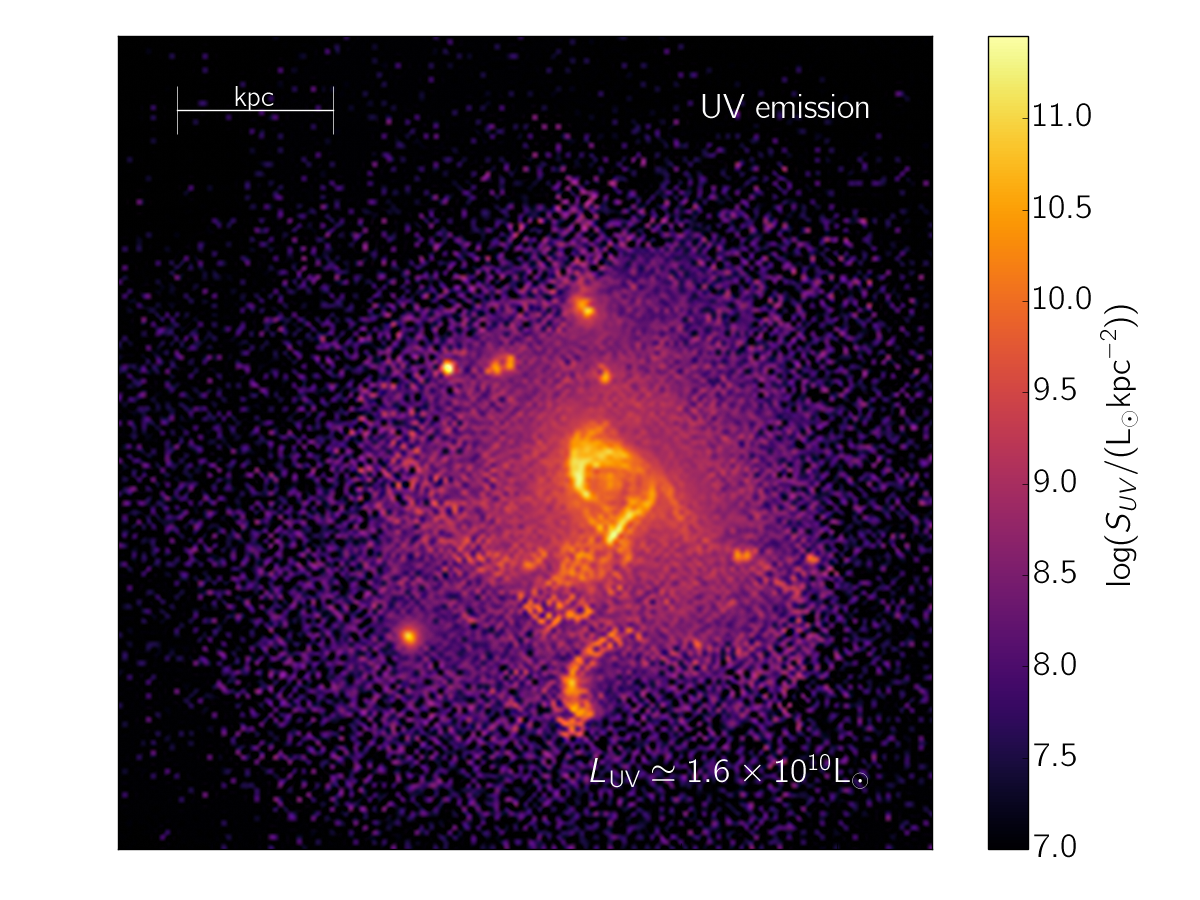}~
\includegraphics[width=0.49\textwidth]{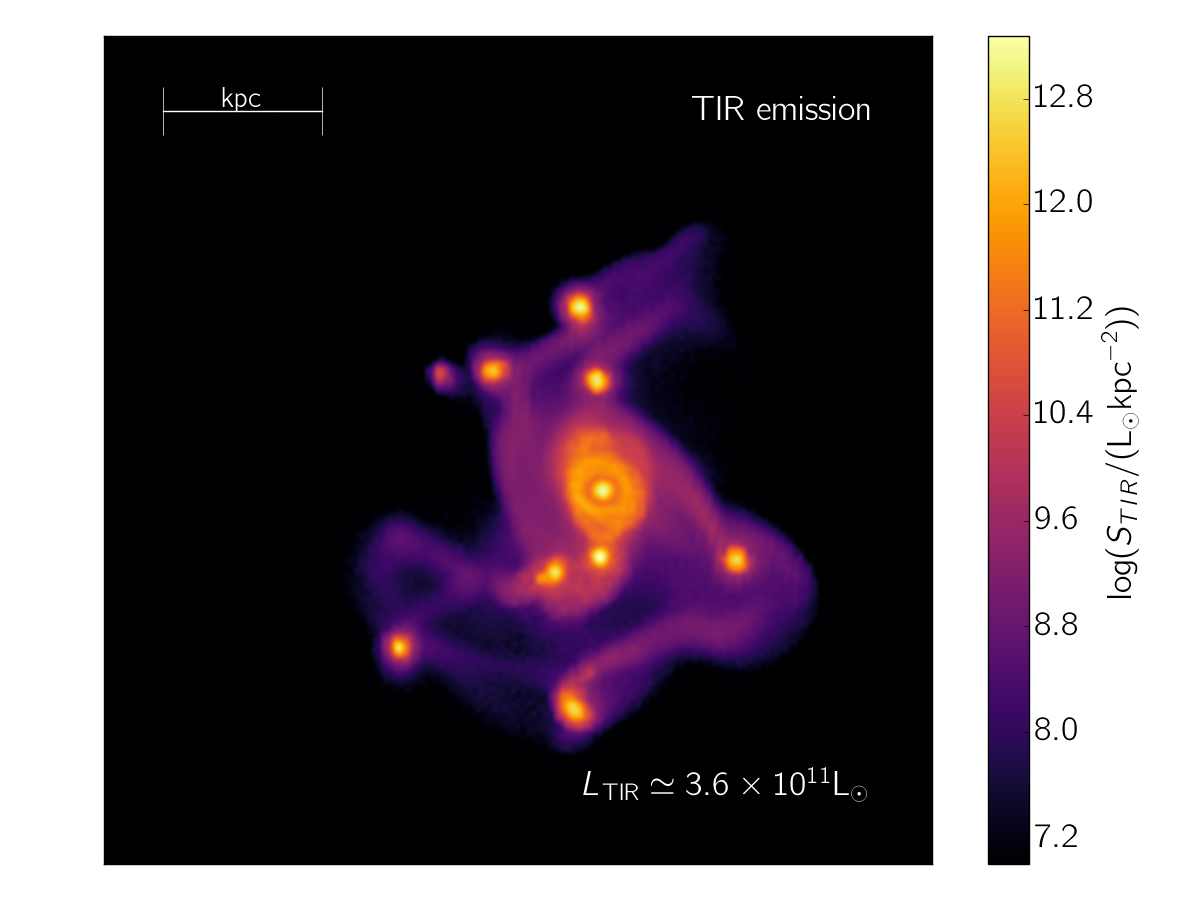}
\caption{
Face-on map of Althaea showing UV (left) and total IR (TIR, right) emission for our fiducial run. With respect to Fig. \ref{fig_althaea_ph}, the maps are centered/zoomed on the peak of UV emission, but have the same orientation.
\label{fig_map_uv_and_tir}
}
\end{figure*}

The result from our fiducial radiative transfer simulation can be appreciated in Fig. \ref{fig_map_uv_and_tir}, where we show the the UV (from 0.1 to 0.3 $\mum$) and total IR (TIR, evaluated from 8 to 1000 $\mum$) emission arising from the galaxy, seen in a face-on orientation.
Both bands are dominated by the luminous disk of the galaxy. The disk has an effective radius $r_e=0.46$ kpc, is largely molecular and is responsible for most of the [CII] emission (\P17cit).
Note that the UV emission is more extended than the FIR, consistent with the fact that a fraction of UV photons is scattered at locations far away from their origin.

On top of the disk, several small knots connected by a filamentary structure are recognizable. 
Such knots are linked to the clumpy structure of the molecular gas in Alth{\ae}a (see Fig. \ref{fig_althaea_ph} and Fig. 6 in \P17cit). Their size is about 50 pc, with total gas masses of $\sim 10^{6-7} \msun$, of which $\simeq 10\%$ is molecular\footnote{We recall that in Alth{\ae}a the maximum molecular fraction of the gas is about $10\%$. This is caused by the chemistry modelling combined with the high interstellar radiation field (ISRF) and the $30$ pc resolution.}; additionally the clumps are rich in metals, with $Z\simeq \zsun$ and in some cases even slightly higher.
Both the central clump and the knots are sites of very recent star formation, with young stars boosting the UV.
At the same time, these small pockets also contain a relatively large amount of dust, in turn leading to strong UV absorption and re-emission in the FIR. Yet, the knots are visible in both the UV and IR; star formation inside of them is only partly obscured. For the interpretation of these results, it is crucial to keep in mind that the pockets provide the majority of the IR flux, while the UV radiation is mostly coming from region where the attenuation is not that strong, as detailed later in the work.

\subsubsection{Dust optical depth}

\begin{figure}
\centering
\includegraphics[width=0.49\textwidth]{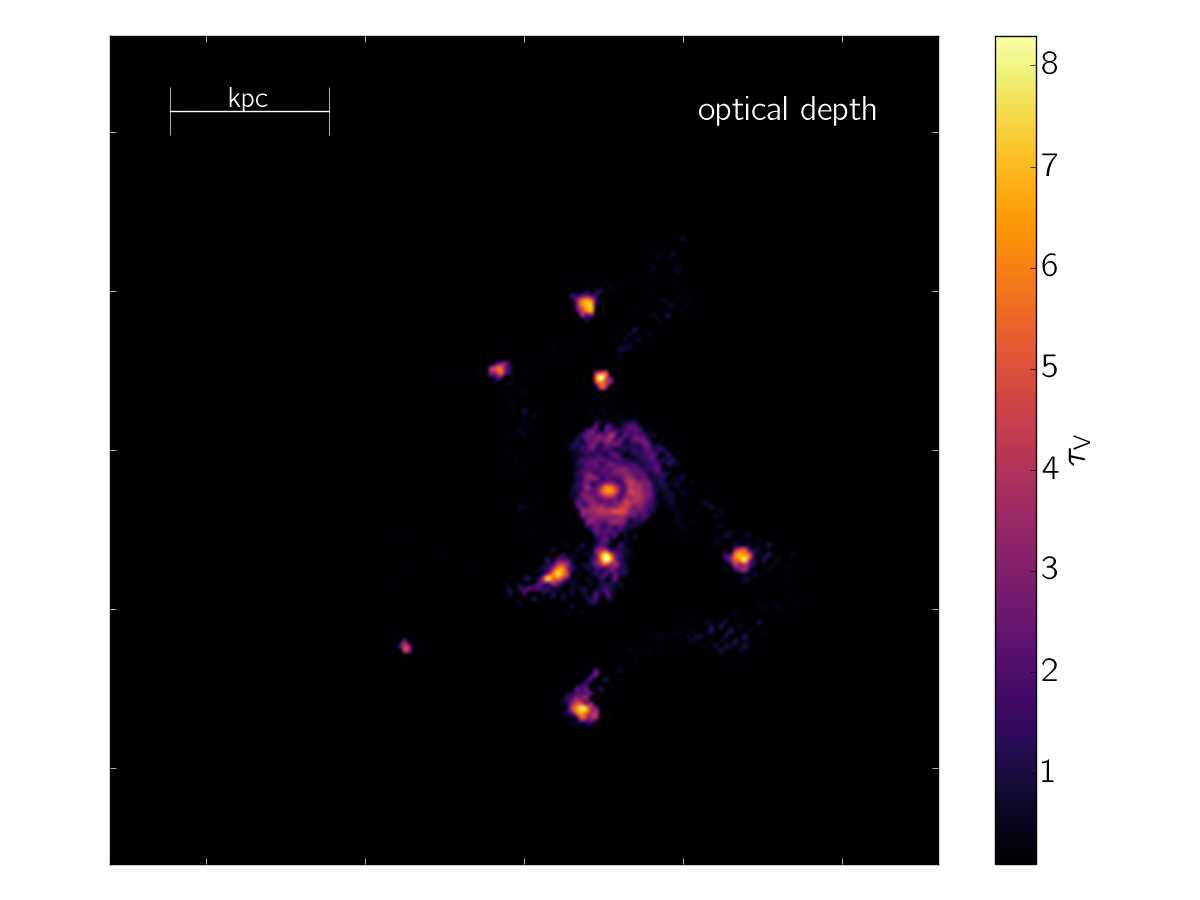}
\caption{
Map of the optical depth $\tau_V$ evaluated at $\lambda = 0.551 \mu$m. The field of view is the same as in Fig. \ref{fig_map_uv_and_tir}.
\label{fig_map_tau}
}
\end{figure}

The attenuation of stellar emission is typically parametrized by $A_V$, the total visual attenuation, which is identical to the optical depth $\tau_V$ up to a small factor:
\begin{equation}
A_V = 1.086\,\tau_V\,.
\end{equation}
This quantity can be inferred from observed spectra; without spatially resolved data, each observed galaxy is associated with a single value for the optical depth, that is a weighted mean of the $\tau_V$ values of the small-scale structure of the observed object.

For Alth{\ae}a (fiducial run) such weighted value, computed as explained below, is $A_V=1.6$, i.e. about two times larger than reported by \L17cit. 
From our simulation, we can investigate the origin of the larger resulting attenuation in more detail. We compute $\tau_V$ by taking the log of the ratio of the intrinsic, $F_{V}^{0}$, to the processed, $F_{V}$, flux:
\begin{equation}
\tau_V = - \ln (F_{V}/F_{V}^{0})\,.
\end{equation}
We find $\tau_V=1.4$ for our fiducial case. We stress that this value represents an \textit{effective} optical depth, i.e. $\tau_V$ is the result of the combined transmission from a large number of lines of sight (los, $\mathbf{x}$) through the body of the galaxy:
\begin{equation}
F_{V} = \sum_{los} F_{V}^0(\mathbf{x}) \exp(-\tau_{V}(\mathbf{x}))\,.
\end{equation}
In other words, the effective optical depth is given by the intrinsic flux in each los weighted by $e^{-\tau_{v}\mathbf{x}}$, which in general is not equal to the mean optical depth per los:
\begin{equation}
\tau_V = {\rm ln}\left[\frac{\sum_{los} 
F_{V}^0(\mathbf{x}) \exp(-\tau_{V}(\mathbf{x}))}{F_{V}^0}\right] \neq \langle\tau_{V}(\mathbf{x})\rangle\,.
\end{equation}

It is therefore instructive to look at a resolved map of $\tau_V$, shown in Fig. \ref{fig_map_tau}. From the map, $\langle\tau_{V}(\mathbf{x})\rangle = 0.42$ with a large r.m.s. of 0.94. In particular, we see that the optical depth is large at the location of the luminous knots also seen in the FIR and UV maps (see Fig \ref{fig_map_uv_and_tir}), with $\tau_V$ up to 8. As reported in Sec. \ref{sec_overview_emission}, these knots are sites of recent star formation, with relatively young stars enshrouded in dust. We find that about 80\% of the total transmission in the V band comes from lines of sight with  $\tau_V < 2$. On the other hand, about 85\% of the TIR emission comes from sites with $\tau_V > 2$, namely the star forming knots. This dichotomy has important implications for recovering the physical properties of such an object, as we will discuss in Sec. \ref{sec_sed_fitting}.

\subsection{Dust temperature}\label{sec_temperature}
The SED of the fiducial model (Fig. \ref{fig_sed_panels}, middle panel) has the peak of IR emission at relatively short wavelengths (400 $\mum$, or about 50 $\mum$ restframe). Consequently, we expect at least a fraction of the dust in Alth{\ae}a to be very warm.

\begin{figure}
\centering
\includegraphics[width=0.49\textwidth]{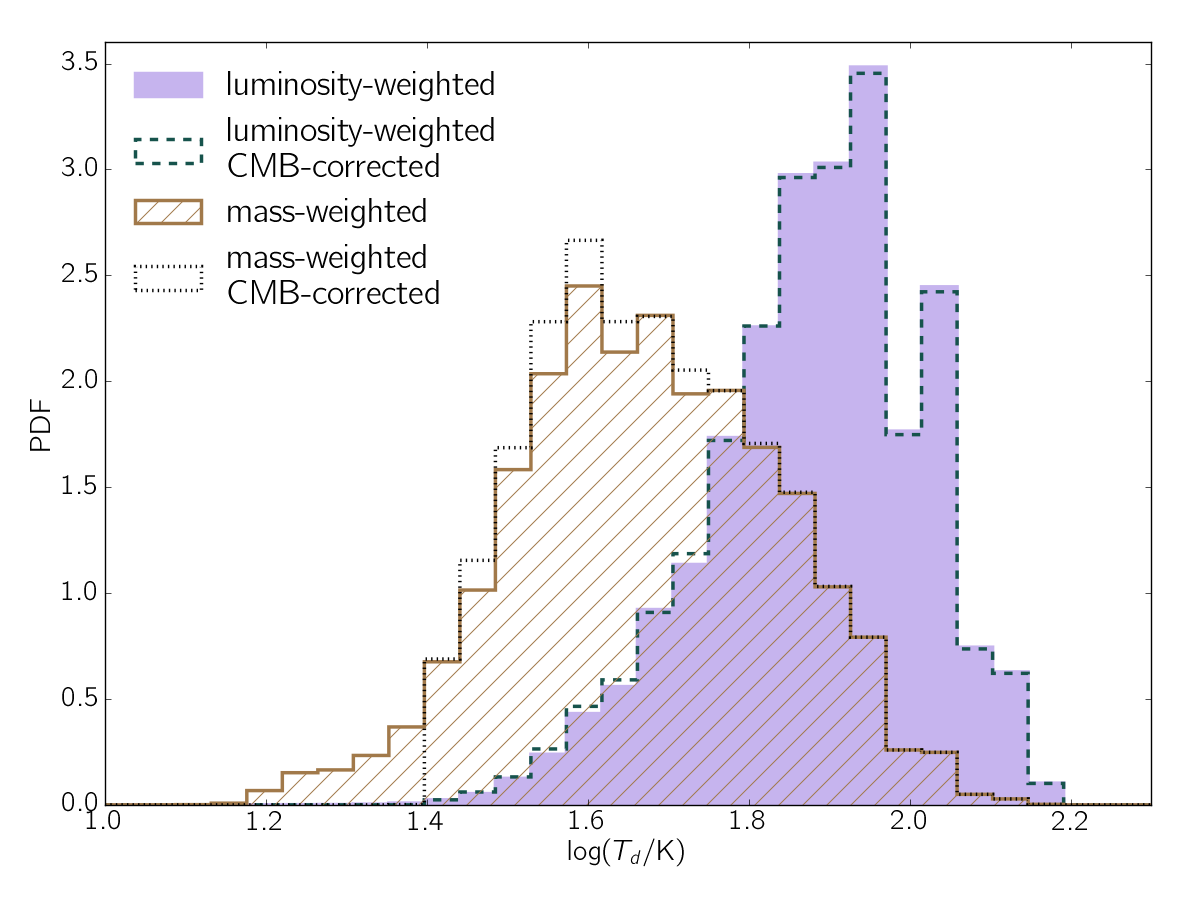}
\caption{
Distribution of dust temperatures in the fiducial model. We show both the mass-and luminosity-weighted ($L\propto M_{D} T_{D}^{4+\beta_d}$) dust temperature, and with/without accounting for CMB heating.
\label{fig_pdf_temp}
}
\end{figure}

In Fig. \ref{fig_pdf_temp}, we show the probability distribution function (PDF) of dust temperatures ($T_D$) for our fiducial run.
Hotter dust does not only emit at shorter wavelengths, but entails a higher luminosity; thus it is important to distinguish between the mass-weighted and luminosity-weighted temperature PDF, i.e. assuming for simplicity a grey body radiation with $L \propto M_D T_D^{4+\beta_d}$ for each cell, with $\beta_d=1.7$ as obtained from the SED.
In general, both weighting schemes lead to single-peaked PDFs, with a similar spread of $\sim$ 20 K. Weighting by luminosity suppresses the low-temperature tail and shifts the peak to higher temperatures, while the mass-weighted is more symmetric.
The two weighting schemes yield significantly different average dust temperatures, because a relatively small mass fraction of high $T_D$ dust dominates the total dust emission. More precisely, we find that only $25\%$ of the dust mass has $T_D > 60$ K, but that fraction contributes $80\%$ of the total luminosity.
Since what is typically observed is a measure of the IR luminosity, the luminosity-weighted scheme should be more comparable with observations. In this case, we find an average temperature of $91 \pm 23 \,\rm K$, which is relatively warm compared to e.g. \L17cit, i.e. $\simeq 50\,\rm K$ (see Tab. \ref{tab_magphys}). While our mass-weighted mean ($50 \pm 19 \,\rm K$) is closer to the value reported by \L17cit, we stress that the correct comparison is the one with the luminosity-weighted value, since the value obtained by \L17cit comes from the SED fitting process and is luminosity-weighted, \citep[see][eq. 8]{daCunha2015}.

As we will argue in detail in Sec. \ref{sec_sed_fitting}, the uncertainty in estimating the dust temperature will be large if it relies on a single FIR data point. For example \L17cit adopted a modified black body for their SED-based estimate\footnote{we note that this estimate done by \L17cit is independent of full SED fitting they performed, see Sec. \ref{sec_sed_fitting}}, assuming $T_D$ temperature to be in a range from $37-55\,\rm K$, and fit the ALMA data point alone.

As we have already seen from the emission  (Fig. \ref{fig_map_uv_and_tir}) and optical depth (Fig. \ref{fig_map_tau}) maps, young clusters of stars are embedded in dense pockets of dust. The compactness of these regions leads to a strong interstellar radiation field, i.e. with ISRF of $G\simeq 100\, G_{0}$, with $G_{0}=1.6\times 10^{-3} {\rm erg}\,{\rm cm}^{-2}\,{\rm s}^{-1}$ being the average MW value for the FUV flux in the Habing band ($6-13.6\,{\rm eV}$ \citealt{habing:1968}).

Since the optical depths in these regions are high, dust can be heated up to large temperatures. Note that $\tau_V$ in these clouds is large even for low dust-to-metal ratios, $f_\mathrm{d}=0.08$ (see \ref{fig_map_tau}). That means that the vast majority of the UV photons is absorbed, namely a fraction $f = 1-e^{-\tau_{V}} \approx 99\%$. If we increase the dust mass, the optical depth increases linearly, but $f$ remains about constant. That means that in the range explored for $f_\mathrm{d}$, the total luminosity absorbed by dust is nearly independent of the dust mass in the clumps. However, the absorbed luminosity is distributed onto a larger or smaller amount of dust in each region. Consequently, increasing $f_\mathrm{d}$ (and therefore $\tau_V$) reduces the resulting average temperature of the dust, and the IR SED is shifted to longer wavelengths (upper and middle panel of Fig. \ref{fig_sed_panels}).

Finally, we also show the temperature distribution of dust including the correction for the additional heating from the CMB in Fig \ref{fig_pdf_temp}. The major effect of this correction is that the minimum $T_D$ is set by the CMB; this can be easily seen for the mass-weighted distribution. However, in the luminosity-weighted case, the bulk of the PDF is already above the CMB temperature without the correction, so the CMB has only a very marginal effect. This implies that the SED is only marginally affected by the CMB correction as well.

\subsection{The IRX-$\beta$ relation}
\begin{figure*}
\centering
\includegraphics[width=0.98\textwidth]{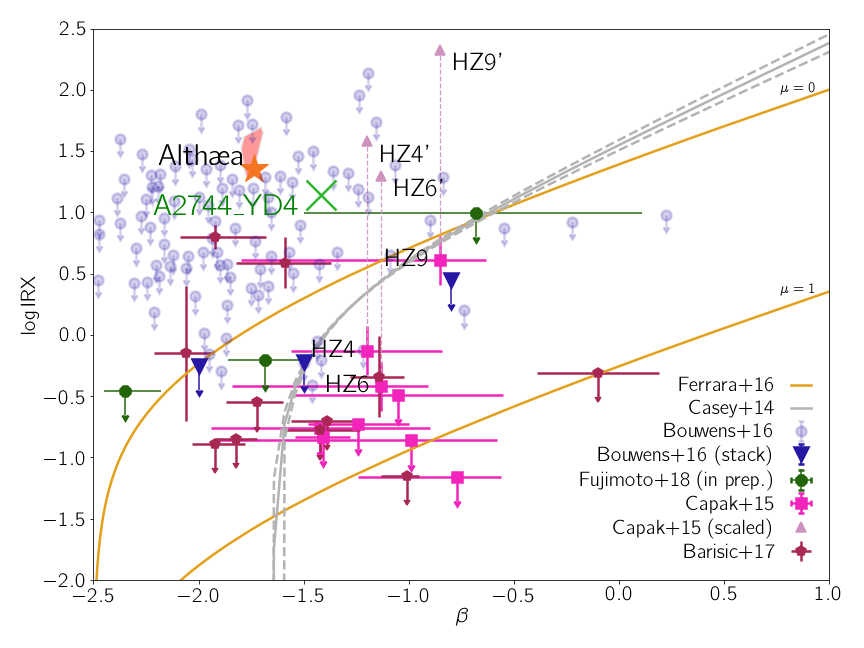}
\caption{
IRX-$\beta$ relation. We plot the result for our fiducial model for Alth{\ae}a (orange star) along with the value inferred from the observation of A2744\_YD4 (green cross). The gray/red region above the star shows the spread of IRX/$\beta$ when considering different viewing angles in our simulations (see App. \ref{appendix_inclination}). Additionally, we plot data for local galaxy \citep[][grey lines]{Casey2014}, various high-z upper limits and detections collected by \citet[][blue circles and triangles]{Bouwens2016}, \citet[][violet squares]{Capak2015}, and model for high-$z$ galaxies \citep[][orange lines]{Ferrara2016} as function of their molecular gas fraction $\mu$.
\citet[][]{Capak2015} galaxies are also plotted by rescaling the inferred dust temperature (pink triangles) and as re-analysed by \citet[][red pentagons]{Barisic2017arXiv}.
See the text for the details and the definitions given in eq. \ref{eq_def_irx} and \ref{eq_def_beta}.
\label{fig_irx_beta}
}
\end{figure*}

The relation between the UV-normalized \quotes{infrared excess} (IRX) and the spectral UV slope (defined as $\propto \lambda^\beta$) is a frequently used diagnostic for both local and high-redshift galaxies \citep{Meurer1995,Calzetti2000,Takeuchi2012,Reddy2012,AlvarezMarquez2016}. The IRX \footnote{Note that observationally, the full SED is usually not accessible, and the denominator of eq. \ref{eq_def_irx} is evaluated as $L_\mathrm{UV} \approx L_{0.16 \mum} \times 0.16 \mum$. 
We have checked that the resulting IRX values from using the latter definition differ only at the percent level.} is defined as follows:  
\begin{equation}\label{eq_def_irx}
\mathrm{IRX} = \frac{\int_{8\;\mu \mathrm{m}}^{10^3\;\mum}  L_\lambda d\lambda}{\int_{0}^{0.3\;\mum}  L_\lambda d\lambda}\, \simeq \frac{L_\mathrm{TIR}}{L_\mathrm{UV}}\,.
\end{equation}
The observed UV slope $\beta$ is defined as
\begin{equation}
\beta = { \frac{\log F_{\lambda_1} - \log F_{\lambda_2}}{\log {\lambda_1} - \log {\lambda_2}} }\,,
\label{eq_def_beta}
\end{equation}
where $(\lambda_1, \lambda_2) = (1600, 2500)$ \AA  (restframe). 

From the theoretical point of view, as extinction typically decreases with wavelength, the UV slope becomes flatter when the attenuation/dust mass is increased. At the same time, the increased energy absorbed in the UV boosts the IR emission. Thus, on general grounds, a positive correlation between the two quantities is expected \citep[see][for an in-depth discussion]{Ferrara2016}.

In Fig. \ref{fig_irx_beta}, we show the location of Alth{\ae}a in the IRX-$\beta$ plane (star symbol). A cross indicates the position of A2744\_YD4; for this calculation, the best fit model that \L17cit recovered using SED fitting was used.

We compare these results with the fitted IRX-$\beta$ relation for local galaxies \citep{Casey2014} and observations and upper limits for high-redshift objects from \cite{Bouwens2016}, \cite{Capak2015}, and \citet{Barisic2017arXiv}. Note that \cite{Bouwens2016} data are all upper limits and the value obtained from stacking is also shown; we also show stacking results from Fujimoto(in prep.), which is a reanalysis of the Bouwens data.
We also report model predictions of \citet[][]{Ferrara2016}, with ($\mu = 1$) and without ($\mu = 0$) molecular gas.

We find that both Alth{\ae}a and A2744\_YD4 have a significantly larger IRX than the observational samples of \citet[][$z=5-6$]{Capak2015} and Fujimoto(in prep.)($z>5$) given their UV slope. While the upper limits of \cite{Bouwens2016} seem compatible, this agreement vanishes when stacking the data. Comparing with the predictions of the \citet[][]{Ferrara2016} model, we find a mismatch even for galaxy without molecular gas ($\mu=0.0$). The crucial point here is that \citet[][]{Ferrara2016} considers that the high dust density in molecular clouds shields the dust from being heated, and thereby can reduce the IR flux. The assumption entails such clouds to be externally heated, whereas we find young stars embedded in the clumps, that can be very efficiently heated from the inside.

To further understand the differences with respect to the observational sample, it is useful to first compare the UV and IR luminosities.
In the \citet[][]{Capak2015} sample, all except for one galaxy  have a higher UV luminosity than Alth{\ae}a or A2744\_YD4 ($\log(L_{\rm UV}/\lsun) \approx 10.2$). This is not surprising since the sample consists of galaxies with a low dust content, and consequently low attenuation. The IR luminosity, on the other hand, shows a dichotomy. While in \citet[][]{Capak2015} $>50$\% of galaxies have $\log (L_{\rm TIR}/\lsun) < 10.5$, about 10 times fainter than Alth{\ae}a/A2744\_YD4, three objects (HZ4, HZ6, and HZ9)\footnote{We note that also HZ10 has a large IR luminosity, but only a upper limit for the UV slope. We therefore omit HZ10 here.} have IR luminosities between $11 < \log (L_{\rm IR}/\lsun) < 12$. These objects are characterized by large SFRs ($>50 \msunyr$), similar to Alth{\ae}a. However, at face value, these galaxies have relatively large UV fluxes compared to Alth{\ae}a and thus, a lower IRX.

The IR luminosity for the \citet[][]{Capak2015} sample has been estimated by fitting modified gray bodies to the (single) FIR data point/upper limits for each object. This involves an assumption concerning the dust temperature. \citet[][]{Capak2015} varied $T_D$ in the range of (25, 45) K. This is substantially lower compared to what we find in the simulation. If indeed the dust in early galaxies is much hotter, as we suggest here, the TIR flux should be highly enhanced. To illustrate this point, we show the FIR-detected \citet[][]{Capak2015} objects after scaling the data points up to a fixed temperature of 90 K (same symbols without black edges, labelled HZ4'/6'/9'), i.e. we multiplied the IR luminosity by $(90 {\rm K}/45 {\rm K})^{4+\beta_d}$. As discussed before, this increases IRX substantially. Although rough, this  procedure serves to elucidate the crucial dependence of the IRX on the assumed dust temperature.

Complementary to this argument, our result is also more consistent with what has been proposed recently by \citet{Barisic2017arXiv}, who re-analysed the \citet{Capak2015} sample using new HST/WFC3 near-infrared imaging for a more accurate determination of $\beta$. They find an excess in the IRX-$\beta$ relation in some objects, and advocate that these galaxies have substantially altered spatial dust distributions, as we indeed find for Alth{\ae}a.

As a matter of fact, our result can be interpreted in terms of the two component emission from Alth{\ae}a, that is given by the sum of dusty IR-bright clumps and a UV-bright smooth component. While the IRX is set by the combination of both components, the $\beta$ slope is dominated by the smooth, UV-bright component; in fact, the slope of the smooth component is -1.8, close to the total value (-1.7, also see Fig. \ref{fig_sed_tau}), while its corresponding IRX is 0.5, in better agreement with the expectations from the local IRX-$\beta$ relation. This explanation is also consistent with the finding of \cite{Casey2014} that higher redshift galaxies tend to have higher IRX values with respect to their $\beta$ slopes (e.g. see their figure 3, middle panel). While \cite{Casey2014} consider only redshifts up to $\sim$ 3, we speculate that this trend extends to the redshifts considered here. 

Additionally, our finding of a higher IRX with respect to the UV slope is consistent with \cite{Narayanan2017}; they report to find higher IRX relative to $\beta$ for shallower extinction curves like the MW curve we use. A shallower extinction tilts the intrinsic UV SED only weakly for relatively high attenuation, shifting the galaxy to the left in the IRX-$\beta$ plane. 

Further, in Fig. \ref{fig_irx_beta} we show the scatter of IRX/$\beta$ due to different inclinations with the gray/red shaded region. The scatter is small in both variables. Our fiducial orientation (face-on) tends to have a relatively lower value for IRX compared to other orientations; this happens because the IR flux does not change with viewing angle, while the UV tends to be suppressed as we move closer to edge-on directions (see App. \ref{appendix_inclination} for details).

Finally, we note that the excess IRX that we find propagates to other diagnostics as well, e.g. to the ratio of CII and TIR luminosity, $L_{CII}/L_{TIR}$, versus star formation rate surface density ($\dot{\Sigma}_{\star}$). We compare here our simulation with the relation found by \citet[][]{smith2017apj}. Alth{\ae}a has a mean $\dot{\Sigma}_{\star} = 1.44\,\msun/{\rm pc}^2/\myr$, and we find that the IR emission is $L_{TIR} = 3.6 \times 10^{11} \lsun$ (Fig. \ref{fig_map_uv_and_tir}); calculating the [CII] emission according to the \citet{vallini:2015} model we have $L_{CII} = 6.8 \times 10^{7} \lsun$ (see \citealt[][]{Pallottini2016}; \P17cit for details of the calculation). Thus, for Alth{\ae}a we have $L_{CII}/L_{TIR} \simeq 0.02\%$, while \citet[][see in particular Fig. 3]{smith2017apj} report a value of $\simeq 0.5\%$ for galaxies with the same $\dot{\Sigma}_{\star}$ (i.e. similar to HZ10 with values from \citealt{Capak2015}). Given the discrepancy in the IRX-$\beta$ diagnostic, this is qualitatively not surprising, and can be attributed again to the high dust temperature of our galaxy. However, the discrepancy in $L_{CII}/L_{TIR}$ is less pronounced.

%
\section{Recovering physical  parameters from SEDs}\label{sec_sed_fitting}
\begin{figure}
\centering
\includegraphics[width=0.49\textwidth]{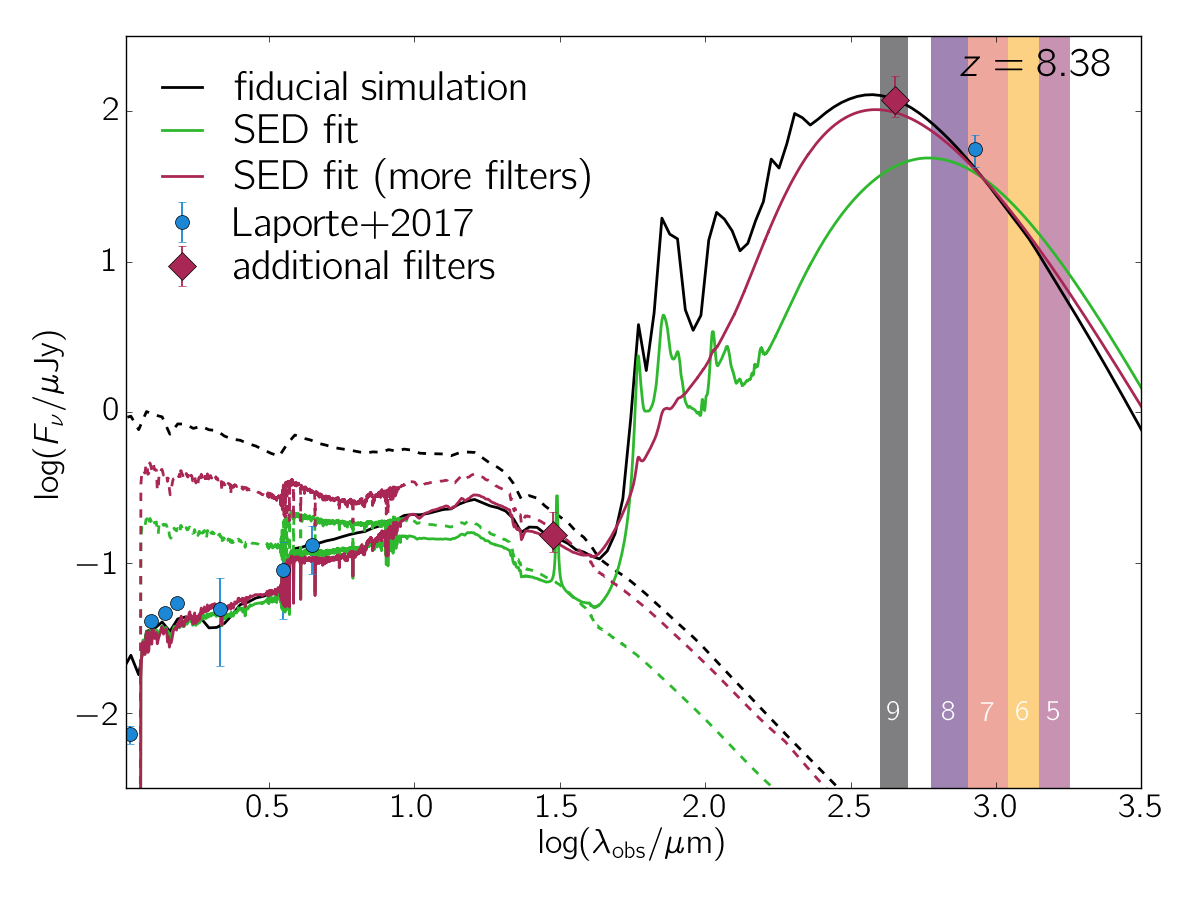}
\caption{
Comparison of the intrinsic/processed SED from our fiducial run (solid/dashed, black) and the intrinsic/processed best-fit model recovered by \code{Magphys} \citep{daCunha2008,daCunha2015} using the same filters as \protect\L17cit (solid/dashed, green). Additionally, we show the intrinsic/processed, recovered SED  using two more filters (indicated with red symbols) in mid/far-IR. Physical properties from the simulation and as recovered by the SED fit are reported in Tab. \ref{tab_magphys}.
\label{fig_sed_final}
}
\end{figure}

\begin{figure}
\centering
\includegraphics[width=0.49\textwidth]{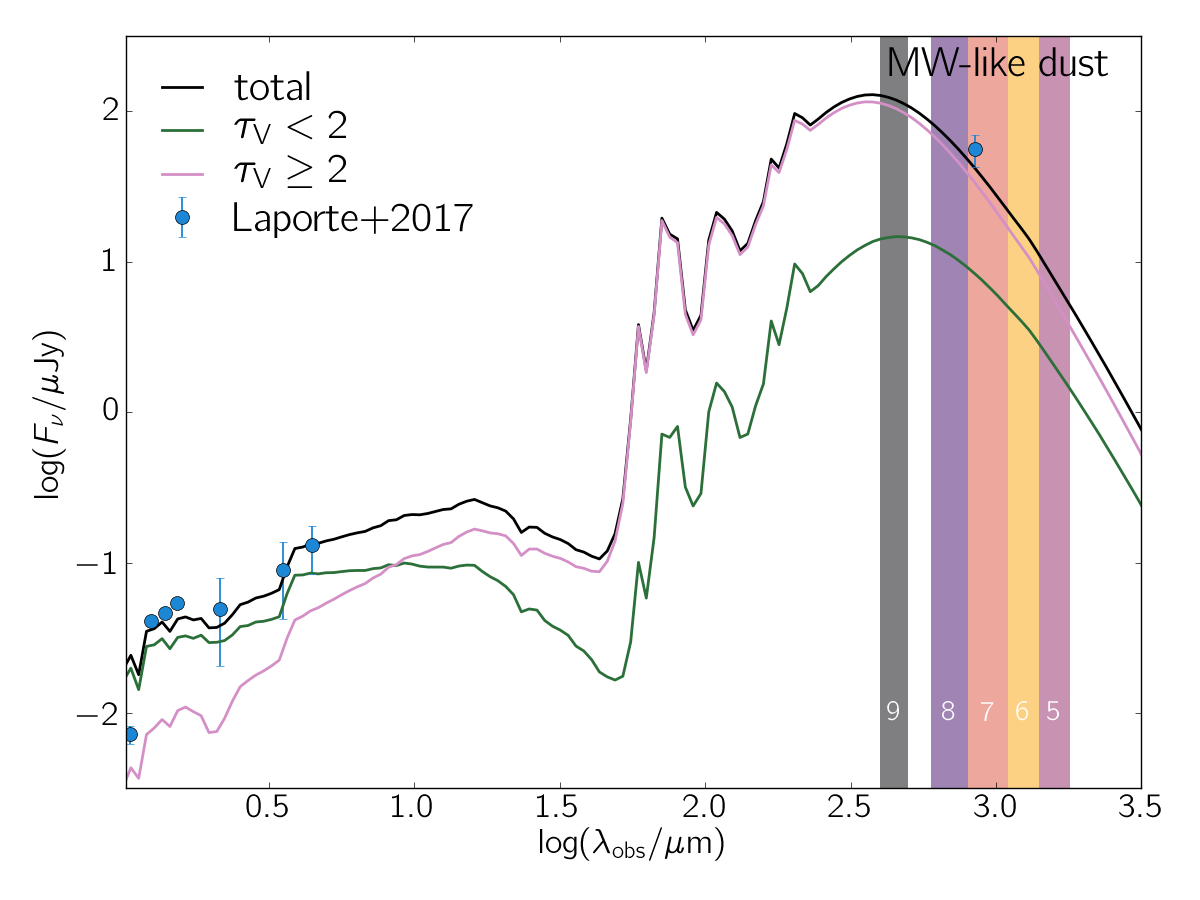}
\caption{
Fiducial simulation SED for the two components: the smooth component (green, $\tau<2$) dominates the UV, the clumpy component (red, $\tau\geq 2$) contributes to most of the IR flux, because of the to high attenuation. For comparison, the total SED (black) is also shown, together with the data points from \protect\citep{Laporte2017} .
\label{fig_sed_tau}
}
\end{figure}

\L17cit used the publicly available \code{Magphys} code \protect\citep{daCunha2008,daCunha2015} to recover physical properties from their observed SED data points. \code{Magphys} uses a sophisticated model for stellar emission, attenuation, and reemission in the IR, and fits both the stellar-dominated/dust-dominated part of the SED. Emission and reemission are connected by energy conservation: up to some tolerance, the energy loss from attenuation in the UV contributes to the total IR energy. By fitting to a large number of template SEDs, \code{Magphys} builds the likelihood distribution of physical parameters like the stellar mass, the SFR, or the dust temperature.

Having a simulation at hand that fits the \L17cit data allows us to compare the physical parameters recovered by the SED fitting to the known values from Alth{\ae}a. We use the high-$z$ version of the code, and the filters available in \L17cit observations. The resulting best-fit model from \code{Magphys} (solid, green) is compared with the SED from our fiducial simulation (solid, black) in Fig. \ref{fig_sed_final}.

As is evident from the plot, the recovered SED deviates from the original one significantly in the mid and far IR.
This is expected since we have only one data point in the IR (ALMA Band 7): with these conditions the IR-part of the spectra is only loosely constrained, because by changing the total energy in the IR and the characteristic dust temperature, the IR spectrum can be shifted and tilted in different ways, all of which fit the IR data. In our case the recovered SED from \code{Magphys} has its peak in the IR at much larger wavelengths than in the original SED, corresponding to dust with relatively low temperature ($T_D = 53 \pm 11 \, \rm K$).
We stress that since \code{Magphys} checks all the models in the library to find the likelihood of physical parameters, this is not problematic per se: quantities constrained by the IR part of the SED simply will be recovered with larger uncertainties. In our case, however, the recovered parameters are far off the true parameters when taking into account the reported uncertainties (see Tab. \ref{tab_magphys}).

To illustrate this point, we ran \code{Magphys} again, including two additional filters in the IR (indicated with red diamonds) to constrain the IR part of the SED. The recovered best-fit model is shown in Fig. \ref{fig_sed_final} (solid, red) and, as expected, is closer to our simulated SED, with a higher dust temperature of $T_D = 61 \pm 6\,\rm  K$. It is worth noting here that \code{Magphys} parametrizes dust emission by considering both dust in thermal equilibrium and stochastically heated dust.

Stochastically heated dust is modeled using two components, one component for hot dust, modelled by two modified black bodies at 130/250 K, and one component for PAHs modeled using an empirical template \citep[for details, see][]{daCunha2008}. The equilibrium components are parametrized by temperatures of modified black bodies for the warm/cold dust in the diffuse ISM (T$^{\mathrm{ISM}}_w$/T$^{\mathrm{ISM}}_c$), and a warm dust component in the birth clouds of stars, T$^{\mathrm{BC}}_w$. In the high-redshift version of \code{Magphys}, T$^{\mathrm{ISM}}_w$ is fixed to 45 K, while T$^{\mathrm{ISM}}_c$ can vary from 20-40 K, and T$^{\mathrm{BC}}_w$ can vary from 30-80 K. In contrast, our simulations include the full distribution of grain temperatures. We note that the limited range of temperatures for the diffuse ISM/birth clouds, in particular for the temperature of dust in birth clouds, yields a very asymmetric PDF, making the percentiles harder to interpret. For example, the best fitting model has $T^\mathrm{BC}_w = 79$ K, while the median of the PDF is 61 K as stated above.

In Fig. \ref{fig_sed_final}, we also show the corresponding intrinsic SEDs with dashed lines. Independently of the choice of filters, \code{Magphys} prefers a solution with a lower intrinsic luminosity; while putting stronger constraints on the IR reduces the discrepancy, the recovered rest frame far-UV of the intrinsic spectrum is still lower than the simulated SED by a factor of 5. This is also reflected in the recovered physical parameters; \code{Magphys} underestimates the SFR, and therefore, the factor dominating the far UV luminosity, by a factor of $\sim$ 4 in this case and by a factor of $\sim$ 6 if we use only the filters used by \L17cit. This difference in SFR is much larger than the error bars on the \code{Magphys} values. Similarly, the stellar mass is underestimated by a factor of $\sim$ 5. It compensates for this lack of UV emission by (i) reducing the attenuation ($A_V$ lower by  a factor of 2), (ii) lower dust temperatures, and (iii) significantly higher dust mass. 
In Tab. \ref{tab_magphys}, we summarize the physical properties of our fiducial model and compare them to the ones recovered by \code{Magphys}.

In principle, such bias towards lower SFR can be introduced by the difference in star formation history (SFH) in Alth{\ae}a with respect to what is implemented in \code{Magphys}; while Alth{\ae}a features a SFH that is exponentially increasing with time, in \code{Magphys} the SFH is slowly rising at early times and exponentially decreasing at later times, i.e. ${\rm SFH}\propto \gamma^2 t \exp(-t\gamma)$, with $\gamma$ as a (positive) free parameter of the model.
To assess the situation we used the publicly available code \code{CIGALE}\footnote{version 0.11, \url{http://cigale.lam.fr/}} \citep{burgarella2005,noll:2009,roehlly2014} to fit the simulated SED, by allowing both exponentially increasing and decreasing SFH. While the recovered SFR are indeed higher than the one obtained with \code{Magphys}, they are still a factor $2-3$ lower then the SFR in the simulation. Thus, we conclude that the discrepancy we find can not be understood from the different SFH alone.

However, the discrepancy is likely to originate from the emission dichotomy we find in our simulation, i.e. a clumpy component with large attenuation providing most of the TIR, and a smooth, diffuse component with relatively low optical depth providing most of the UV flux. In Fig. \ref{fig_sed_tau}, we again show the SED from our fiducial simulation (black line), and for comparison the SEDs of the two components by applying a cut at $\tau_\mathrm{V} = 2$. This is illustrating again that the smooth component ($\tau_\mathrm{V}<2$) drives the UV flux, while the clumpy component ($\tau_\mathrm{V}\geq 2$) clearly dominates the IR. In other words, the failure of the SED fitting might be related to the fact that we fit the superposition of two systems, a clumpy, dusty one, and a smooth, less dusty one; the necessity of having two such components has previously been shown, e.g., by \citet{bianchi2008,popescu2011,delooze2012,saftly2015}, and has been suggested for high-$z$ objects by \cite{Barisic2017arXiv}.

Finally, we would like to emphasize the importance of mid-IR data for better constraining the underlying physics. Instruments like the proposed SPICA \citep{spinoglio:2017} will be crucial to understand complex systems, e.g. the composition of their dust grains and the importance of self-absorption (see App. \ref{appendix_convergence}).
\begin{table*}
\centering
\begin{tabular}{llllll}
\hline
~                & $M_{\star}$   & SFR              & $M_{D}$        & $A_V$ & $T_D$ \\
~                & $[10^9\msun]$   & $[\msunyr]$ & $[10^6\msun]$    & ~     & $[\rm K]$ \\  
\hline
\rule{0pt}{1em} A2744\_YD4 \citep{Laporte2017}& $1.97^{+1.45}_{-0.66}$ & $20.4^{+17.6}_{-9.5}$ & $5.5^{+19.6}_{-1.7}$ & $0.74^{+0.17}_{-0.45}$ & 37-63\\ 
\rule{0pt}{1em} fiducial simulation & $10.3$ & $78.0$ & $1.6$ & $1.6$ & $91 \pm 23$\\ 
\hline
Parameters recovered using \code{Magphys} given the fiducial simulation data: & & & & & \\
\rule{0pt}{1em} SED fit & $2.2_{-0.9}^{+0.7}$ & $11.8_{-3.0}^{+3.9}$ & $5.8_{-3.2}^{+13.1}$ & $0.5 \pm 0.2$ & $54_{-12}^{+10}$\\ 
\rule{0pt}{1em} SED fit (more filters) & $2.8_{-0.8}^{+0.8}$ & $16.7_{-2.6}^{+4.8}$ & $4.1_{-1.7}^{+8.7}$ & $0.7_{-0.2}^{+0.2}$ & $62_{-9}^{+8}$\\
\hline
\end{tabular}
\caption{Summary of physical properties of the simulated galaxy Alth{\ae}a in the fiducial \code{SKIRT} run and as recovered by Magphys for the stellar mass ($M_{\star}$), star formation rate (SFR), dust mass ($M_{D}$), extinction ($A_V$) and dust temperature ($T_D$). Note that the error for $T_D$ for Alth{\ae}a is the r.m.s. of the dust temperature. See Fig. \ref{fig_sed_final} for the corresponding SED. For comparison, we also quote the physical parameters estimated by \protect\L17cit in the first row of the table.
\label{tab_magphys}
}
\end{table*}

\section{Discussion and conclusions}\label{sec_conclusions}

We have used the dust radiative transfer code \code{SKIRT} to post-process Alth{\ae}a, the zoom-in simulation of a prototypical LBG in the epoch of reionization.
We compared the resulting stellar and dust continuum fluxes to the observations of A2744\_YD4 \citep{Laporte2017}, a recently discovered dusty galaxy at $z=8.38$. We caution the reader that Alth{\ae}a has not been devised to be a constrained simulation of the object observed in \citet{Laporte2017}, however we can make an educated guess for a direct comparison by using an evolutionary state of Alth{\ae}a that matches A2744\_YD4 by redshift, stellar mass (up to a factor of 2) and star formation rate. While we do not find a match for this specific evolutionary stage, we were able to reproduce the observed SED by varying the dust-to-metal fraction and the intrinsic UV emissivity (i.e. star formation rate).
By simultaneously matching the UV and FIR data we obtain the physical properties of the galaxy (for a summary see Tab. \ref{tab_magphys}). The key findings are:
\begin{itemize}
\item[\bf a)] The star formation rate of A2744\_YD4 is ${\rm SFR} = 78\, \msunyr$, i.e. a factor $\approx 4$ higher compared to the Spectral Energy Distribution fitting employed by \cite{Laporte2017}.
\item[\bf b)] Dust attenuation is consistent with a Milky Way extinction curve and we find a mean optical depth $\tau_V=1.4$.
\item[\bf c)] The dust-to-metal ratio is low, $f_\mathrm{d} \sim 0.08$, implying that early dust formation is rather inefficient (or that dust destruction is very efficient).
\item[\bf d)] The luminosity-weighted dust temperature is high, $T_d=91\pm 23 \,\rm K$, as a result of the intense interstellar radiation field ($\approx 100\times$ MW).
\item[\bf e)] Due to the high $T_d$ the ALMA Band 7 detection can be explained by a limited dust mass, $M_d=1.6\times 10^6 M_\odot$.
\item[\bf f)] Higher dust temperatures might solve the puzzling low infrared excess recently deduced for high-$z$ galaxies from the IRX-$\beta$ relation.
\end{itemize}
  The dust emission in our simulation is dominated by the several star forming knots within the disk of Alth{\ae}a. These regions are characterized by large dust column densities, with $\tau_V$ up to 8, and large SFR surface densities (mean $\dot{\Sigma}_{\star} = 1.44\,\msun/{\rm pc}^2/\myr$). Consequently, the ISRF at these sites is about a factor of 100 larger than in the MW, leading to efficient heating of dust grains. Together with the small value for $f_\mathrm{d}$, this results in dust grain temperatures of around 90 K when using the luminosity-weighted average, with a small fraction of dust at large temperatures contributing the vast majority of IR radiation. On the other hand, the UV emission from the galaxy is dominated by regions of lower optical depth ($\tau_\mathrm{V}<2.0$). This can be interpreted as Alth{\ae}a being composed of two different components, the clumpy, IR-bright one with high attenuation, and the smooth, diffuse that is UV-bright.

Note, however, that our model does not include photoevaporation and/or full radiative feedback, which might affect the structure of the simulated galaxy. 

To investigate the mismatch between the physical parameters found by \L17cit and the ones we need to employ to fit their observational data, we perform the same SED fitting technique as these authors. The SED fitting cannot recover the real properties of our simulation, and we find that it tends to prefer low attenuation, low SFR solutions, even when we increase the number of data points fed to the SED fit.
This may be related to the fact that the employed code assumes SFR histories of the form $\gamma^2\exp(-\gamma t)$, whereas Alth{\ae}a has an exponentially increasing SFR history. However, comparison with a different SED fitting tool (\code{Cigale}) does not resolve this issue, although it allows for a different SFR history.
We suggest that the mismatch is driven by assumption about the temperature of the dust in the employed models, preferring lower temperature than what we find from the full simulations, and by the fact that Alth{\ae}a consists of two components with very different properties. Therefore, our simulations suggest that the physical properties of A2744\_YD4 may be significantly different from the parameter estimation in the literature, that is, it might be characterized by a large SFR around 80 $\msunyr$ and a smaller dust mass of $10^6\msun$, with a significant fraction of the dust heated to high temperatures, boosting the IR.

In the IRX-$\beta$ diagnostics, we find Alth{\ae}a and the SED fit of \L17cit to A2744\_YD4 to lie above the typical relation for local galaxies and also above the observed high-redshift galaxies, that is, both galaxies are brighter in the IR than implied by the their $\beta$ values.
However, the derivation of IRX for high-redshift galaxies typically requires an assumption for the dust temperature, and we find much higher temperatures in our simulation ($\sim 90$ K) compared to what is assumed ($< 60$ K).
If we correct for this, we find higher and more comparable IRX values for the observed objects, e.g. for the \citet{Capak2015} sample. We also note that \cite{Barisic2017arXiv} partly find significantly lower values for $\beta$ for objects from the \cite{Capak2015} sample, moving them closer to Alth{\ae}a and the \L17cit galaxy. Additional care has to be taken since the assumed model for the dust grains can significantly alter the value of the UV slope $\beta$.
Finally, we note that $\beta$ is dominated by the UV-bright component, whereas IRX results from the combination of the clumpy, IR-bright component and the UV-bright component, which leads to high values for IRX relative to $\beta$; this is consistent with the findings of \cite{Casey2014}, although they consider lower redshift galaxies.

In order to match \L17cit data, we had to set the dust-to-metal fraction $f_\mathrm{d}$ to a lower value than deduced for the MW or the SMC.
Taken at face value, we can use this result to constrain the dust yield at high redshifts.

As in Alth{\ae}a we know the stellar age of each star particle, we can calculate the dust yields by distinguishing between the production from Supernovae and Asymptotic Giant Branch (AGB) stars using the same model adopted in the hydrodynamical simulation (see Sec. \ref{sec_hydro_sim}); this is possible by assuming the dust production to be proportional to the metal yield, i.e. a fixed dust condensation fraction is taken for both Supernovae and AGB.
Then, in our fiducial run we find that in Alth{\ae}a about 10\% of the total metal mass is produced by AGB stars, and 90\% by Supernovae; Supernovae are dominant since stars that are old enough to be in the AGB phase account only for $\simeq 5\%$ of the mass, as a result of Alth{\ae}a formation history.

The value of $f_d$ needed to fit the data now translates into a dust yield of $2 \times 10^{-3}$ M$_\odot$ per stellar mass, meaning that either dust production is inefficient, or processes destroying dust are efficient enough to reduce the dust mass significantly.
However in the current simulation we do not follow the evolution of the dust properties, thus at the present we cannot distinguish between these processes and further we note that mass losses from AGB stars -- and thus yields -- are very uncertain in stellar modeling \citep[e.g.][]{ventura:2005,riebel:2012}; this is an interesting aspect worth investigating in the future. In perspective, a large statistical sample of simulations like Alth{\ae}a will be necessary for a statistical assessment. While is this is out of reach for the present, our proof-of-concept work here shows that such a comparison can yield useful insights on the properties of high-redshift galaxies.

\section*{Acknowledgements}
We are grateful to Peter Camps for code support, interesting discussion and the development of \code{SKIRT}.
We thank Ricardo Amorin, Ilse De Looze, Hiroyuki Hirashita, Roberto Maiolino for helpful discussion, and Nicolas Laporte for providing data.
AF acknowledges support from the ERC Advanced Grant INTERSTELLAR H2020/740120
We acknowledge use of the Python programming language \citep{VanRossum1991}, and use of the Numpy \citep{VanDerWalt2011}, IPython \citep{Perez2007}, Jupyter \citep{LoizidesSchmidt2016}, Matplotlib \citep{Hunter2007}, and Pymses \citep{Labadens2012} packages. This research made use of Astropy, a community-developed core Python package for Astronomy \citep{astropy}.

\bibliographystyle{mnras}
\bibliography{paper,packages} 



\appendix

\section{Numerical convergence and additional physical processes}\label{appendix_convergence}

\begin{figure}
\centering
\includegraphics[width=0.49\textwidth]{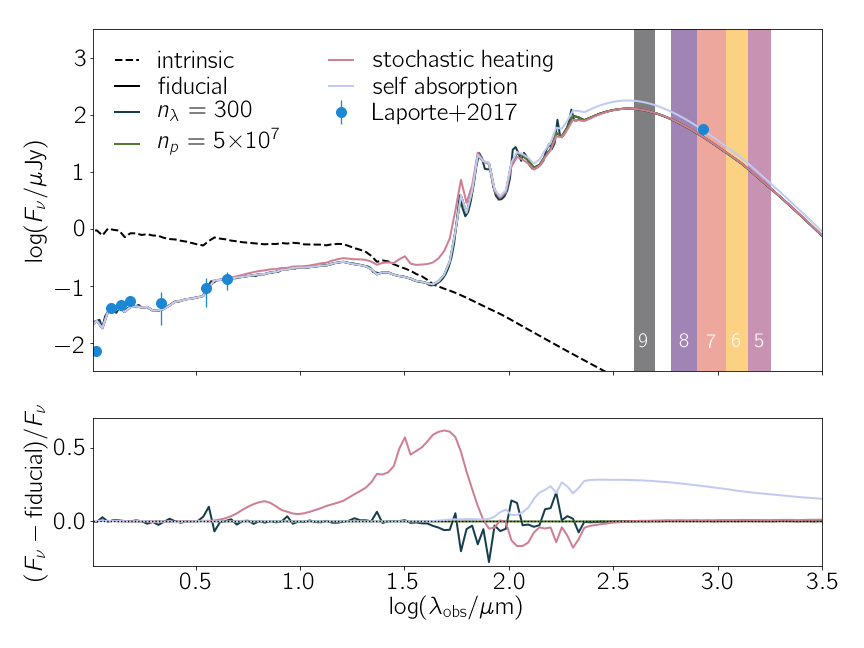}
\caption{In the {upper panel} we plot the SED for the fiducial case and two runs with more wavelength bins ($n_{\lambda}=300$) and more photon packages ($n_{p}=5\times 10^7$) launched per wavelength bin. Additionally, we show the outcome of a simulation for which stochastic heating was turned on. In the {lower panel} we show the fractional difference between the various runs and the fiducial model
\label{fig_convergence}
}
\end{figure}

In this appendix, we check the numerical convergence of our radiative transfer simulations.
The two most relevant parameters in \code{SKIRT} are the number of photon packages launched per wavelength grid point, $n_p$, and the number of grid points used for the wavelength grid, $n_\lambda$.
By default, we used ($n_p$,$n_\lambda$) = (10$^7$, 150). In Fig. \ref{fig_convergence}, we show the SEDs for $n_\lambda=300$ and $n_p = 5 \times 10^7$. The differences are small ($< 5\%$) in the region where the \L17cit data lies, in particular are negligible when integrated over the observed filters, thus do not influence our results.

Additionally, in Fig. \ref{fig_convergence} we also report a case in which we used stochastic heating, that is, we dropped the assumption that all grains are in thermal equilibrium with the incident radiation field \citep[see][for a detailed analysis of such effect]{camps:2015_heat}. The stochastic heating run results in an increased emission at about 30-80 $\mum$. While the differences in the mid-IR are significant (up to $\simeq 50\%$), both the UV and the FIR show little change, thus leaving our results unchanged. However, we note that such mid-IR window well overlaps with the spectral range of the proposed instrument SPICA \citep{spinoglio:2017,egami:2017}, that thus would be a crucial probe in distinguishing the effect of stochastic heating for these high-$z$ galaxies.

As a last test, we have considered self-absorption of dust. Similarly to our previous test, we reran fiducial simulation with self-absorption switched on \citep{Baes2011}. The resulting SED is shown in Fig. \ref{fig_convergence}, and it features a net decrease of flux in the mid-IR and an increase in the FIR, stemming from the self-absorption of dust-emitted radiation by dust grains, and the consecutive reemission. The maximum FIR flux is about 40\% higher when we include self-absorption. While this effect is not negligible, including it does not change our interpretation significantly. A value of $f_\mathrm{d}\sim 0.08$ remains the preferred one, mainly because the UV is unchanged by allowing self-absorption. Self-absorption affects the temperature of dust grains, but in our case, the change is about 3\% with respect to the fiducial simulation. We stress, however, that self-absorption has an non-negligible effect on both the mid-IR and the FIR, and should be considered carefully. Similar to the case of stochastic heating, we note that the variation in the mid-IR could potentially be used to gauge the impact of self-absorption, using instruments like SPICA.

Finally, since we do not let \code{SKIRT} reconstruct the dust distribution, but instead use exactly the grid structure of the hydro simulation, there is no need to test whether the sampling of the dust distribution is converged.

\section{The effect of inclination}\label{appendix_inclination}

\begin{figure}
\centering
\includegraphics[width=0.49\textwidth]{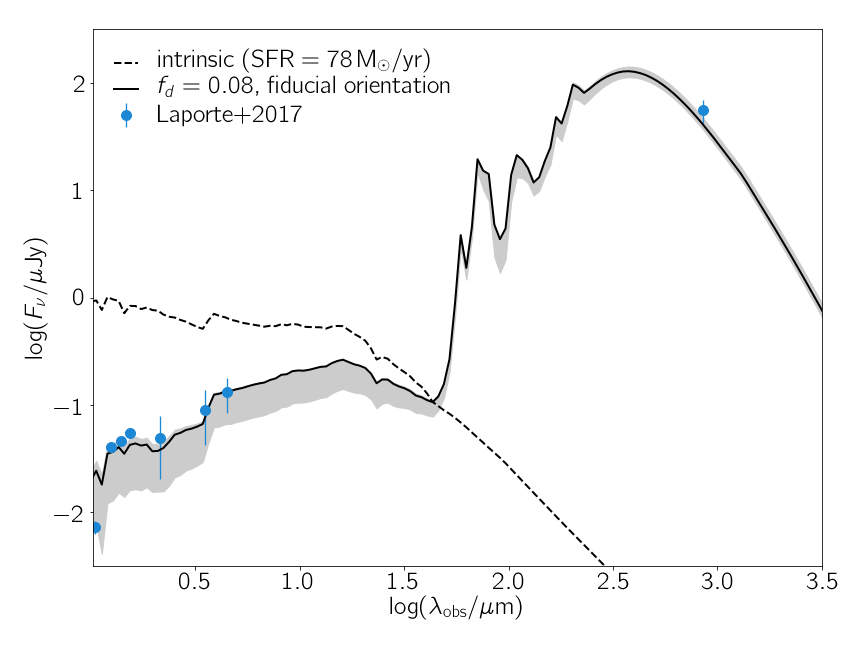}
\caption{Comparison between our fiducial model/orientation (solid black line) with 48 other lines of sight, homogeneously distributed on the sphere. The latter are shown in terms of their spread (gray contour).}
\label{fig_inclination}
\end{figure}

Since different lines of sight might result in different observed SEDs, we reran our fiducial simulation with 48 lines of sight, covering the sphere using an equal area and iso-latitude tessellation \citep{Healpy}. We show the result in Fig. \ref{fig_inclination}. We find that the FIR fluctuates only marginally, since it escapes isotropically from the simulated galaxy\footnote{We recall that dust self-absorption is considered only in App. \ref{appendix_convergence}}. However, the UV is subject to absorption that is stronger in edge-on directions, thus the spread is larger in the UV. Since our fiducial line of sight is face-on, it has large UV fluxes compared to other orientations. For example, the flux in the J band at about 1.2 $\mu$m has a value of $(0.026\pm 0.007)\,\mu{\rm Jy}$ within the sample, i.e. it shows a variation of $\simeq 27\%$ depending on the line of sight. The fiducial line of sight has a higher flux of 0.032 $\mu{\rm Jy}$, about 20\% higher than the average, as expected for a face-on direction. Fitting the data to strongly attenuated edge-on directions would require a lower dust attenuation (lower $f_\mathrm{d}$) and a higher intrinsic UV flux, i.e. a higher SFR, because lowering $f_\mathrm{d}$ would reduce the FIR flux, which can only be recovered by an higher intrinsic UV flux.

\bsp    
\label{lastpage}
\end{document}